\documentclass[prd,aps,10pt,twocolumn,floats,floatfix,nofootinbib] {revtex4-1}

\usepackage{graphicx}
\usepackage{dcolumn}
\usepackage{bm}
\usepackage{graphics}
\usepackage{slashed}
\usepackage{amsfonts}
\usepackage{amssymb}
\usepackage{natbib}
\usepackage{amsmath}
\usepackage{url}
\usepackage[usenames,dvipsnames,svgnames,table]{xcolor}
\usepackage{color}
\usepackage{xcolor}
\usepackage{tabularx}
\usepackage{comment}
\usepackage{hyperref}
\usepackage{enumitem}
\usepackage{soul}

\hypersetup{
    colorlinks=true,
    linkcolor=red,
    citecolor=blue,
}

\begin{document}
\title{Bekenstein bounds in maximally symmetric nonlinear electrodynamics}
\author{Juan M. Diaz${}^{1}$}\email{juan-manuel.diaz@etu.univ-amu.fr}
\author{Marcelo E. Rubio$^{2,3,4}$}\email{marcelo.rubio@gssi.it}
\affiliation{\vspace{0.2cm}
${}^{1}$Aix Marseille University, 3 Pl. Victor Hugo, 13003 Marseille, France\\
%${}^{2}$Instituto de F\'{\i}sica Enrique Gaviola (CONICET), Medina Allende y Haya de la Torre, s/n, Córdoba, Argentina\\
${}^{2}$Gran Sasso Science Institute (GSSI), Viale Francesco Crispi 7, I-67100 L’Aquila, Italy\\
${}^{3}$INFN, Laboratori Nazionali del Gran Sasso, I-67100 Assergi, Italy\\
${}^{4}$SISSA, Via Bonomea 265, 34136 Trieste, Italy
}

\begin{abstract}

We explore dynamical features of the maximally symmetric nonlinear extension of classical electromagnetism, recently proposed in the literature as ``ModMax'' electrodynamics. This family of theories is the only one that preserves all the symmetries of Maxwell's theory, having applications in the study of regular black hole solutions and supersymmetry. 
The purpose of this article is three-fold. Firstly, we study the initial-value problem of ModMax and show, by means of a simple geometric criterion, that such a theory admits a well-posed formulation. Secondly, we prove a series of geometric inequalities relating energy, charge, angular momentum, and size in ModMax. The validity of these bounds gives strong evidence of an universal inequality conjectured by Bekenstein for macroscopic systems. Finally, we perform the first stable numerical simulations of ModMax in the highly nonlinear regime. We apply our simulations for verifying an inequality between energy, size and angular momentum in bounded domains, and for showing birefringence as a function of the nonlinear parameter, comparing with other nonlinear nonlinear extensions.
\end{abstract}

\maketitle
%\tableofcontents

%%%%%%%%%%%%%%%%%%%%%%%%%%%%%%%%%%%%%%%%%%

\section{Introduction}
\label{sec-Intro}

The first proposal of a nonlinear theory for electromagnetism came up in the early 30's, developed by Born and Infeld with the aim to resolve the singularity of the electric field produced by a point charge \citep{Born-Infeld1934}. Since then, nonlinear extensions of classical electrodynamics have attracted attention, akin their applications to a wide range of phenomena. Spanning from the theory of fundamental interactions to gravitational and cosmological scenarios \citep{beatoPRL98,Cataldo:2000ns,Yajima:2000kw,ELIZALDE20031,PhysRevD.69.127301}, nonlinear electromagnetic effects also take shape in condensed matter systems \citep{Hartnoll:2009sz,Zaanen:2015oix}, dielectric crystals \citep{PhysRevB.4.3694}, and nonlinear optics-- the vacuum birefringence being one of their main warhorses \citep{PhysRevLett.125.221301,Greco:2022ufo}. Some of these models are also useful for the description of conducting materials using holographic arguments \citep{Lai:2021mxa,JING201068}.

Although most of the proposals for nonlinear electrodynamics (NLED) are framed as effective field theories for electromagnetic interactions, for some of them the corresponding fundamental theories are well known. For instance, the seminal theories proposed by Born-Infeld \citep{Born-Infeld1934} and Euler-Heisenberg \citep{Heisenberg:1936nmg} are based, respectively, on String Theory and Quantum Electrodynamics. The latter, in particular, accounts for nonlinear corrections to the classical theory due to quantum electron-positron interactions at one loop. Interestingly, some other extensions arise upon dimensional reduction of a certain higher-curvature gravity theory, as the one proposed by Kaluza and Klein for describing the low-energy limit of String Theory \citep{Kaluza:1921tu,PhysRevD.63.064006}.

In General Relativity, gravitational collapse ends up in black-hole configurations which inevitably have curvature singularities, and the field equations are no longer valid \citep{PhysRevLett.14.57}. Such equations should then be replaced by some quantum theory or effective modified gravitational theory that explains the dynamics of the spacetime surrounding the singularity. Remarkably, by coupling NLED to gravity, it is possible to construct stable and regular black-hole solutions, thus avoiding the development of physical singularities \citep{PhysRevD.106.024031,Franzin:2021vnj,PhysRevD.103.084052,Simpson:2018tsi}.

From a general perspective, nonlinear extensions of electromagnetism are supposed to generalize Maxwell's classical theory, for it is natural to wonder which of them (if existing) preserve its fundamental symmetries. It is widely known that the classical theory is \textit{conformally invariant}, but also invariant under the so-called \textit{duality rotations} (in four spacetime dimensions) \citep{Jackson:1998nia}. Very recently, a family of nonlinear theories has been proposed, which preserve both of the aforementioned symmetries. Moreover, it is the only one sharing this extraordinary property. This family of theories, known as ``ModMax'' (which stands for ``Modified Maxwell''), was proposed in 2020 by Bandos et al., as a conformal limit of Born-Infeld generalizations in String Theory \citep{ModMax2020}. Indeed, ModMax constitutes a weak field limit of a generalized two-parameter Born-Infeld field theory \citep{Kruglov:2021bhs}. 

Although it remains still not clear whether or not ModMax can be consistently quantized, or whether it can be obtained as a conformal limit of a theory with matter-interacting fields which are effectively integrated out, several interesting features arise with it. For example, ModMax admits \textit{ birefringence} (that is, a characteristic structure of multiple light cones), plane wave solutions, and even topologically nontrivial knotted-null configurations, derived from the Hamiltonian formalism \citep{Babaei-Aghbolagh:2024uqp,Kuzenko:2024zra,Babaei-Aghbolagh:2022itg}. This effect admits limited and quite indirect experimental evidence in almost any nonlinear extension of Electromagnetism, as reported by certain nonlinear optics experiments \citep{PVLAS,OSQAR}. Nonetheless, it has many theoretical motivations, coming from quantum gravity, string theory and super-symmetry. For instance, plane-wave solutions in ModMax are useful for resolving apparent shortcomings involving null fields in the theory. This theory also admits supersymmetric extensions up to six dimensions, as well as superconformal higher-spin generalizations \citep{Bandos:2021rqy,Ferko:2024zth,PhysRevD.75.027502,Perry:1996mk}. In \citep{Barrientos:2024umq,Barrientos:2022bzm}, nontrivial black holes sourced by ModMax Electrodynamics were explored, in particular reporting on the first accelerating black hole ever found in a NLED and the first black hole solution embedded on an exterior electromagnetic field, also in a NLED. Remarkably, it has been studied how the intrinsic non-linearity of ModMax alters the interactions between electrically and magnetically charged point particles. In fact, it has recently been shown that the Lienard-Wiechert fields are exact solutions of the ModMax dynamical equations, unlike most NLED candidates \citep{PhysRevD.106.016009}.

In this article, we inspect several features about the dynamics of ModMax, in particular through \textit{geometric inequalities} between energy, charge, angular momentum and size in bounded regions of space. Geometric inequalities relate magnitudes that have both a geometrical meaning and a physical interpretation and have a very long history in mathematics. In General Relativity, the Kerr-Newman family of black hole solutions satisfy important inequalities bounding their mass, charge and angular momentum, even in dynamical regimes. They have been extensively studied in a series of works \cite{DGenRG(1),DGenRG(2),DGenRG(3),DGenRG(4),DGenRG(5),DGenRG(6)}, highlighting their role in gravitational collapse, as well as in the validity of the cosmic censorship conjecture. The soundness of such inequalities does not often depend on the model, although for astrophysical systems they are not simple to display given their complicated internal structure.

An enlightening example is the inequality
\begin{equation}
\label{eq:DesigualdadOriginal}
\frac{\hbar c}{2 \pi k_B} S \leq \mathcal{E}\mathcal{R},
\end{equation}
which was conjectured by Bekenstein \cite{Bekenstein1981} and expected to hold for arbitrary macroscopic configurations. Here, $S$ and $\mathcal{E}$ are respectively the entropy and the energy of the system; $\mathcal{R}$ is the radius of the smallest sphere containing the whole system; $\hbar$ is the reduced Plank's constant, $c$ is the speed of light in vacuum and $k_B$ is the Boltzmann's constant. Inequality (\ref{eq:DesigualdadOriginal}) was soon generalized by Hod, Bekenstein and Mayo \citep{Hod1,Hod2,Bekenstein1981} to include electric charge $Q$ and angular momentum $\mathcal{J}$, giving rise to the improved bound
\begin{equation}
\label{eq:DesigualdadGeneral}
\frac{\hbar c}{2 \pi k_B} S \leq \sqrt{(\mathcal{E}\mathcal{R})^2 - c^2 \mathcal{J}^2} - \frac{Q^2}{8 \pi}.
\end{equation}
By assuming that the entropy must be always non negative, inequality (\ref{eq:DesigualdadGeneral}) implies the following lower bound for the energy:
\begin{equation}
\label{eq:DesigualdadSinS}
\frac{Q^4}{64 {\pi}^2 \mathcal{R}^2}+\frac{c^2 \mathcal{J}^2}{\mathcal{R}^2} \leq \mathcal{E}^2.
\end{equation}
For non-rotating systems, inequality (\ref{eq:DesigualdadSinS}) reduces to
\begin{equation}
\label{eq:DesigualdadSinJ}
\frac{Q^2}{8 \pi \mathcal{R}} \leq \mathcal{E},
\end{equation}
while in electro-vacuum it yields
\begin{equation}
\label{eq:DesigualdadSinQ}
\frac{c \vert \mathcal{J}\vert}{\mathcal{R}} \leq \mathcal{E}.
\end{equation}

Bekenstein and Mayo conjectured that inequalities (\ref{eq:DesigualdadSinS}),(\ref{eq:DesigualdadSinJ}) and (\ref{eq:DesigualdadSinQ}) have universal validity, and therefore they should give a reasonable description of the dynamics of the system, even in highly-nonlinear regimes. In fact, they are valid in classical electromagnetism \cite{Dain2015}, as well as in Born-Infeld theory \cite{Penafiel2017}. Here we show that they also hold in ModMax, giving a strong evidence of the soundness of such universal conjecture.

An essential stride for assessing the consistency of ModMax trough the aforementioned inequalities, is a detailed study of the well-posedness of the corresponding Cauchy Problem. We refer a system of equations as \textit{well-posed} \cite{Hadamard1902} if, for any given initial-data set, (i) there locally exists a solution; (ii) such a solution is unique; and (iii) the solution depends continuously on the initial data, according to the topological spaces on which both the data and the solution belong to. This fundamental condition, proposed by Hadamard in the 20's, is of utmost importance to ensure the predictability of any physical theory, as well as asymptotic decay of the solution, providing estimates on the existence time, among other aspects \cite{NH(1),NH(2),NH(3),Reula-Abalos20,KreissOrtiz2014,Reula-Rubio2017}. In this work, we prove that ModMax equations admit a well-posed initial-value problem, giving rise to the feasibility of stable numerical simulations.

\subsection{Outline and conventions}

The outline of the paper is the following. In Section \ref{sec-ModMaxPrelim}, we review preliminary aspects on nonlinear extensions of electromagnetism, and revisit the highlights of ModMax theory, fixing the notation that will be used throughout the work. We also introduce some definitions and discuss a few results concerning well-posedness in NLED, particularly through the concept of hyperbolicity. In Section \ref{sec-Wellposedness}, we make use of a simple criterion to show that ModMax is symmetric-hyperbolic, implying a well-posed Cauchy problem and ensuring the feasibility of stable numerical simulations. Section \ref{sec-BekensteinBounds} is devoted to prove the validity of a series of inequalities relating energy, charge, angular momentum and size in ModMax. In Section \ref{sec-Numerics} we describe our numerical set-up used for simulating slab-symmetric field configurations. After showing the dynamical evolution of ModMax in the high nonlinear regime, we extract from the simulations the birefringence effects and compare it with other nonlinear extensions of electromagnetism. Convergence tests are also displayed and, as an application of our simulations, we numerically verify the validity of the inequality between energy, angular momentum and size proved in Section \ref{sec-BekensteinBounds}. Finally, an overall
discussion of our results is left for Section \ref{sec-conclusions}. 

We use the mostly plus signature for the
spacetime metric, $(-,+, +, +)$, and geometric units such that $c=1$ for our simulations.

\section{Preliminaries}
\label{sec-ModMaxPrelim}

To start off, we bring in an overview of nonlinear theories for electromagnetism, introducing some notation and definitions that will be used throughout this work. After particularizing on the maximally-symmetric theory, we elaborate on the corresponding initial-value problem, through the concept of hyperbolicity as an algebraic tool for assessing well-posedness, as well as for characterizing birefringence effects.

\subsection{NLED in a nutshell}

The lagrangian density of relativistic Electrodynamics~\cite{Landau2003}
\begin{equation}\label{eq:LagrangianoMaxwell}
    \mathcal{L}_M=-\frac{1}{4} F_{\mu \nu} F^{\mu \nu},
\end{equation}
is quadratic in the Maxwell tensor $F_{\mu \nu}$, 
which implies that the dynamical equations obtained from varying the corresponding action are \textit{linear} in $F_{\mu\nu}$. However, it is possible to consider corrections to the classical action, therefore extending the electromagnetic theory to \textit{nonlinear} configurations. In fact, consider a general lagrangian density
\begin{equation}\label{general-action}
    \mathcal{L}=\mathcal{L}(F,G),
\end{equation}
as to be a sufficiently smooth function of the electromagnetic invariants
\begin{eqnarray}
\label{Invariantes}
F &:=& \frac{1}{4}F^{\mu \nu}F_{\mu\nu}\,, \nonumber \\
G&:=&\frac{1}{4}\,{}^{*}F^{\mu \nu}F_{\mu \nu}\,, \nonumber
 \end{eqnarray}
where ${}^{*}F^{\mu \nu}= \epsilon^{\mu \nu \sigma \delta} F_{\sigma \delta}/2$ is the Hodge's dual of $F_{\mu\nu}$. The general equations obtained from varying (\ref{general-action}) with respect to $F_{\mu\nu}$ are \cite{Reula-Abalos20}
\begin{eqnarray}
\label{eq:EcENL}
\nabla_{\mu}\left(\mathcal{L}_F F^{\mu \nu} + \mathcal{L}_G {}^{*}F^{\mu \nu} \right)&=&0 \\
\label{eq:EcENL2}
\nabla_{[\mu} F_{\nu \sigma]}&=&0\,,
 \end{eqnarray}	
 where $\mathcal{L}_X$ denotes partial derivative with respect to $X$. Maxwell's equations are recovered by setting $\mathcal{L}=\mathcal{L}_M$ in the above equations. Dynamical evolution is usually given in terms of the electric field $E^\mu=-F^\mu{}_\nu t^\nu$ and the magnetic induction $B^\mu=-{}^{*}F^\mu{}_\nu t^\nu$, measured with respect to an Eulerian observer determined by a timelike, unitary and future-pointing vector field, 
$t^\mu$. Then, the electromagnetic invariants can be rewritten in terms of them as
\begin{eqnarray}
\label{invs-EB}
    F &=& \frac{1}{2}(\mathbf{B}^2-\mathbf{E}^2),\, \nonumber\\
    G&=&\mathbf{E}\cdot\mathbf{B},\nonumber
\end{eqnarray}
where $\mathbf{E}^2=E^\mu E_\mu$ and $\mathbf{B}^2=B^\mu B_\mu$.

Besides the equations of motion, it is useful to characterize NLED theories by a number of quantities obtained directly from the lagrangian density, such as the energy and angular momentum. On a spatial hypersurface $\mathcal{S}$ of the background spacetime, the \textit{electromagnetic energy} contained in a region $\Sigma\subset\mathcal{S}$ is defined by \cite{Penafiel2017}
\begin{equation}
    \label{eq:EnergiaENL}
    \mathcal{E}(\Sigma) = {\int_{\Sigma}^{}\,d^3\Sigma} \left[ -\mathcal{L}_{F}E^\mu E_\mu + 2\mathcal{L}_{G}G - \mathcal{L}\right]. \\
\end{equation}
The \textit{angular momentum} with respect to the center of the smallest sphere containing $\Sigma$, and projected onto an arbitrary direction $\mathbf{k}$ in $\mathcal{S}$, is given by \cite{Penafiel2017}
\begin{equation}
    \label{eq:MomentoAngularENL}
    \mathcal{J}_\mathbf{k}(\Sigma) = -{\int_{{\Sigma}}^{}\,d^3\Sigma}\; \mathcal{L}_{F} (\mathbf{x} \times (\mathbf{E} \times \mathbf{B}))\cdot \mathbf{k}\,, 
\end{equation}
where \textbf{x} is the position of an arbitrary point in $\Sigma$, with respect to the center of the sphere. Lastly, the \textit{electric charge} contained in $\Sigma$ is obtained by integrating the charge density $\rho$, namely
\begin{equation}
\label{eq:CargaENL}
Q(\Sigma) = {\int_{\Sigma}^{}} \, \rho \, .\nonumber
\end{equation}
All these quantities reduce to their familiar expressions in classical electromagnetism. In this work, we will explore the relation between them in a particular family of nonlinear electromagnetic theories, which we describe in what follows.

\subsection{ModMax Electrodynamics}

As it can be easily verified, Maxwell's theory is \textit{conformally} and \textit{dual} invariant \citep{Jackson:1998nia,Wald:1984rg}. The former symmetry means that the lagrangian density is an homogeneous function of first degree; i.e. $\mathcal{L}_M(\varphi F) = \varphi \mathcal{L}_M(F)$, for any smooth scalar field $\varphi$. The latter, instead, means that: if $(F^{\mu\nu},{}^{*}F^{\mu\nu})$ is a solution of Maxwell's equations, then $(\tilde{F}^{\mu\nu},{}^{*}\tilde{F}^{\mu\nu})$ given by the \textit{dual rotation}
\begin{equation}
\label{eq:InvRotDualesMaxwell}
\begin{pmatrix}
\tilde{F}^{\mu\nu} \\
{}^{*}\tilde{F}^{\mu\nu} 
\end{pmatrix}=
\begin{pmatrix}
\cos(\alpha) & \sin(\alpha) \\
-\sin(\alpha) & \cos(\alpha) 
\end{pmatrix}
\begin{pmatrix}
F^{\mu\nu} \\
{}^{*}F^{\mu\nu} \nonumber
\end{pmatrix}
\end{equation}
is also a solution, for any $\alpha\in\mathbb{R}$. It is natural to wonder if there exist \textit{nonlinear} theories that preserve one or both of these symmetries. To answer this, we would first need to generalize the previous definitions, which were stated only for the linear case.

A general theory with lagrangian density $\mathcal{L}(F,G)$ is said to be \textit{conformally invariant} if
\begin{equation}
\label{eq:InvConformeENL}
\mathcal{L}(\varphi F,\varphi G) = \varphi \mathcal{L}(F,G),
\end{equation}
for any smooth scalar field $\varphi$. Also, if $\mathcal{L}$ is smooth enough, a \textit{dual rotation} $(\tilde{F},\tilde{G})$ of a solution $(F,G)$ to Eqs. (\ref{eq:EcENL})-(\ref{eq:EcENL2}) is obtained by the transformation
\begin{equation}
\label{eq:InvRotDualesENL}
\begin{pmatrix}
\frac{\partial\tilde{\mathcal{L}}}{\partial \tilde{F}_{\mu\nu}} \\
{}^{*}\tilde{F}^{\mu\nu} 
\end{pmatrix}=
\begin{pmatrix}
\cos(\alpha) & \sin(\alpha) \\
-\sin(\alpha) & \cos(\alpha) 
\end{pmatrix}
\begin{pmatrix}
\frac{\partial\mathcal{L}}{\partial F_{\mu\nu}} \\
{}^{*}F^{\mu\nu} 
\end{pmatrix},
\end{equation}
where $\tilde{\mathcal{L}}:=4\mathcal{L}(\tilde{F},\tilde{G})$. We say that the theory is \textit{duality invariant} if given $(F,G)$ a solution to (\ref{eq:EcENL})-(\ref{eq:EcENL2}), the pair $(\tilde{F},\tilde{G})$ satisfying (\ref{eq:InvRotDualesENL}) is also a solution, for any value of $\alpha$.

Recently, a remarkable family of nonlinear electromagnetic theories was proposed, known as \textit{maximally-symmetric electrodynamics} (from now on, ``ModMax'', for short) \cite{ModMax2020}. This family is the \textit{only nonlinear extension} of Maxwell's theory which preserves \textit{both} conformal and dual invariance. The corresponding lagrangian density reads 
\begin{equation}
\label{eq:LagrangianoModMax}
\mathcal{L}_\gamma= -\cosh{(\gamma)}\,F + \sinh{(\gamma)}\,\sqrt{F^2+G^2},
\end{equation}
where $\gamma\in\mathbb{R}$ is a dimensionless parameter. The functional form of this Lagrangian is a straightforward consequence of taking a general $\mathcal{L}(F,G)$, and requiring conformal invariance (through the trace of the corresponding energy-momentum tensor) and duality-invariance (which implies a differential equation than can be solved analytically)\footnote{For further details, we refer the reader to the seminal paper, \citep{ModMax2020}.}.

As it can be intermediately noticed, $\mathcal{L}_0 = \mathcal{L}_M$; i.e., Maxwell's theory is recovered by taking $\gamma = 0$. Also, ModMax admits two polarization modes: one which propagates at the speed of light, and other one which is subluminal if $\gamma \geq 0$ and superluminal if $\gamma<0$ \cite{ModMax2020}. Therefore, to ensure causality, we will confine our analysis to the first of these two cases, and take $\gamma\geq 0$. 

The Hamiltonian density of ModMax is obtained, as usual, by a Legendre transformation. In terms of $\mathbf{E}$ and $\mathbf{B}$, it reads 
\begin{equation}
\label{eq:HamiltonianoLegendre}
\mathcal{H}_\gamma=\textbf{E}\cdot\textbf{D}-\mathcal{L}_\gamma,
\end{equation}
where $\textbf{D}$ is the electric displacement vector, i.e. the conjugate variable of $\mathbf{E}$,
\begin{eqnarray}\label{defD}
\textbf{D}&=&\frac{\partial\mathcal{L}_\gamma}{\partial\textbf{E}}.
\end{eqnarray}
We find it useful to express the Hamiltonian in terms of \textbf{D} and $\textbf{B}$, getting
\begin{eqnarray}
\label{eq:HamiltonianoModMax}
\mathcal{H}_\gamma &=& \frac{1}{2}\cosh{(\gamma)}\left(\textbf{D}^2+\textbf{B}^2\right) \nonumber\\
&-& \frac{1}{2}\sinh{(\gamma)}\sqrt{\left(\textbf{D}^2-\textbf{B}^2\right)^2+4\left(\textbf{D}\cdot\textbf{B}\right)^2}.
\end{eqnarray}
Analogously, the conjugate variable of the magnetic induction \textbf{B} is the magnetic field \textbf{H}, given by
\begin{eqnarray}\label{defH}
\textbf{H}&=&-\frac{\partial\mathcal{L}_\gamma}{\partial\textbf{B}}.
\end{eqnarray}
Definitions (\ref{defD}) and (\ref{defH}) correspond to the ``constitutive relations" of the theory. Likewise, it is possible to recover the original variables from their conjugate ones by taking derivatives with respect to the Hamiltonian density; i.e.,
\begin{eqnarray}
\label{eq:EHamiltoniano}
\textbf{E}&=&\frac{\partial\mathcal{H}_\gamma}{\partial\textbf{D}}\,; \nonumber\\
\label{eq:HHamiltoniano}
\textbf{H}&=&\frac{\partial\mathcal{H}_\gamma}{\partial\textbf{B}}.\nonumber
\end{eqnarray}

\subsubsection{Energy density}

The energy density for ModMax is obtained from the corresponding energy-momentum tensor. Recall that for a general nonlinear theory of electromagnetism, the latter is given by \cite{Gibbons2001} 
\begin{equation}
\label{eq:TENL}
T^{\mu \nu}=-\mathcal{L}_F T_M^{\mu \nu} + \frac{1}{4} T g^{\mu \nu},
\end{equation}
where
\begin{equation}
\label{eq:TMaxwell}
T_M^{\mu\nu}={F^\mu}_\rho F^{\nu\rho}-\frac{1}{4}F_{\rho\lambda} F^{\rho\lambda}{g}^{\mu\nu}\nonumber
\end{equation}
is the one corresponding to Maxwell's Electrodynamics. The trace of (\ref{eq:TENL}) is
\begin{equation}
\label{eq:TrazaENL}
T=-4 \left( \mathcal{L}-F \mathcal{L}_F - G \mathcal{L}_G \right),\nonumber
\end{equation}
and from it we can deduce, by explicit computation, that the energy-momentum tensor of ModMax is trace free, i.e., $T=0$. This condition is actually expected from the conformal invariance of the theory. Moreover, ModMax satisfies the \textit{dominant energy condition} (DEC); i.e., for every timelike future directed vector $X^{\mu}$, the inequality 
\begin{equation}
\label{eq:DEC}
\left(T_{\mu \nu} - \frac{1}{2} T g_{\mu \nu}  \right) X^{\mu} X^{\nu} \geq 0,\nonumber
\end{equation}
holds. In fact, the energy density of ModMax is $u=T^{00}$, that is
\begin{equation}
\label{eq:uModMax}
u=\left(\cosh{\gamma}-\sinh{\gamma}\frac{F}{\sqrt{F^2+G^2}}\right)u_M,
\end{equation}
where
\begin{equation}
\label{eq:uMaxwell}
u_M = \frac{1}{2}\left(\mathbf{E}^2+\mathbf{B}^2\right)\nonumber
\end{equation}
is the energy density in classical Electrodynamics. Certainly, as a consequence of the DEC, $u\geq0$ for every value of the fields $\textbf{E}$ and $\textbf{B}$.

The particular case $F=G=0$ deserves a word of caution, as ModMax admits plane-wave configurations \cite{ModMax2020}. Although the energy density (\ref{eq:uModMax}) becomes apparently singular when $F=G=0$, such inconvenience can be removed by means of the Hamiltonian formalism. In ModMax, when both electromagnetic invariants vanish, one gets \cite{ModMax2020}   
\begin{equation}
\mathbf{D}^2+\mathbf{B}^2=2 \cosh(\gamma)\,  \lvert \mathbf{D} \times \mathbf{B}\rvert\,,\nonumber
\end{equation}
which implies that the Hamiltonian \textit{restricted to null fields} is 
\begin{equation}
\mathcal{H}_\gamma\lvert_{\mbox{\tiny{null}}} = \lvert \mathbf{D} \times \mathbf{B}\rvert.\nonumber
\end{equation}
Then, the energy density should be computed from the Hamiltonian, that is
\begin{equation}
u\lvert_{\mbox{\tiny{null}}}=T^{00}\lvert_{\mbox{\tiny{null}}}=\mathcal{H}_\gamma\lvert_{\mbox{\tiny{null}}}, \nonumber    
\end{equation}
which is well defined.

\subsection{Hyperbolicity}

Here we review a few definitions concerning the concept of \textit{hyperbolicity}, as an algebraic tool for assessing the well-posedness of the initial-value problem of any plausible physical theory. These ideas will be essentially used for showing, in a remarkably simple way, that ModMax admits a well-posed Cauchy problem and, moreover, that it is a symmetric-hyperbolic theory, suitable for numerical simulations.

The notion of hyperbolicity captures a key aspect of the evolution equations of any physical system, which should hold even in the most fundamental scenarios. This is a crucial tool for understanding several properties of the dynamical equations, like existence and uniqueness of solutions for a given initial data, preservation of the asymptotic decay of the solution with respect to the asymptotic behaviour of the initial data, and estimates about time of existence of the solutions, among other attributes.

Following \cite{Kreiss97,Kreiss-Ortiz01,kreiss2004initial,geroch1996partial}, and in order to fix notation, we will first consider \textit{linear} first-order systems of partial differential equations; i.e., the ones taking the general form
\begin{equation}
\label{eq:SistemaLineal_0}
\partial_{t} \varphi^{A} +(\mathbb{A}^{A}{}_B(\textbf{x},t))^i \partial_{i}\varphi^B=f^A(\textbf{x},t)\,,
\end{equation}
for a set of dynamical fields $\varphi^A:\mathbb{R}^n \times \mathbb{R}_+ \rightarrow \mathbb{R}^m$, $\varphi^A={(\varphi_1,\ldots,\varphi_m)}^\intercal$ which depend on the spatial coordinates $\textbf{x}={(x_1,\ldots,x_n)}^\intercal$ and time $t$. The functions $f^A:\mathbb{R}^n \times \mathbb{R}_{\geq 0} \rightarrow \mathbb{R}^m$ and the matrices $\mathbb{A}^{j}:\mathbb{R}^n \times \mathbb{R}_{\geq 0} \rightarrow \mathbb{R}^{m\times m}$ depend only on the coordinates and are given. We require the fields to satisfy the initial condition
\begin{equation}
\label{eq:DatoInicial}
\varphi^A(\textbf{x},0)=\varphi_{0}^A(\textbf{x}),
\end{equation}
where $\varphi_0:\mathbb{R}^n \rightarrow \mathbb{R}^m$.
\\
\\
\textbf{Definition I.} System (\ref{eq:SistemaLineal_0})-(\ref{eq:DatoInicial}) is \textit{well-posed} for $t\in[0,T]$ if there exists a unique solution which depends continuously on the initial data. That is, if there exists a norm $\lvert \lvert \cdot \rvert \rvert$ and two real constants $C\geq 0$ and $\beta$ such that, for all $t\in[0,T]$,
\begin{equation}
\label{eq:220}
\lvert \lvert \varphi(\textbf{x},t) \rvert \rvert \leq C e^{\beta t} \lvert \lvert \varphi_{0}(\textbf{x}) \rvert \rvert.\nonumber
\end{equation}
A \textit{purely-algebraic} and \textit{equivalent} condition for guaranteeing well-posedness of the initial-value problem introduced before is that of \textit{hyperbolicity}.
\\
\\
\textbf{Definition II.} System (\ref{eq:SistemaLineal_0})-(\ref{eq:DatoInicial}) is \textit{strongly hyperbolic} if, for every covector $k_i$, the matrix  $A(k):=\mathbb{A}^j k_j$ is diagonalizable and has only real eigenvalues. 
\\

The symbol $\mathbb{A}^j k_j$ is commonly referred to as the \textit{principal symbol} of the system, and contains all the information on the propagation speeds of the theory. A direct consequence of strongly-hyperbolic systems is the existence of a \textit{symmetrizer}; i.e., a positive definite bi-linear form $\mathbb{H}(k)$ such that $\mathbb{H}(k)\mathbb{A}(k)$ is \textit{symmetric} in all its arguments. This motivates the notion of \textit{symmetric-hyperbolic} systems.
\\
\\
\textbf{Definition III.} System (\ref{eq:SistemaLineal_0})-(\ref{eq:DatoInicial}) is \textit{symmetric hyperbolic} if there exists a common symmetrizer $\mathbb{H}$ for all the covectors $k_j$. In other words, the system is symmetric hyperbolic if and only if $\mathbb{H}$ is independent of $k_j$.
\\
\\
The condition for symmetric hyperbolicity is stronger than the one for strong hyperbolicity: symmetric hyperbolic systems are also strongly hyperbolic and, therefore, they constitute a well-posed initial-value formulation \citep{Hilditch:2013sba}. 

All previous definitions can be generalized to \textit{quasi-linear} systems by allowing the principal symbol to depend also on the dynamical fields $\varphi^A$, namely
\begin{equation}
\label{eq:SistemaLineal}
\partial_{t} \varphi^{A} +(\mathbb{A}^{A}{}_B(\varphi,\textbf{x},t))^i \partial_{i}\varphi^B=f^A(\textbf{x},t).
\end{equation}
\textbf{Definition IV.} System (\ref{eq:SistemaLineal}) is \textit{strongly hyperbolic} if there exists a symmetrizer $\mathbb{H}=\mathbb{H}(\textbf{x},t,\varphi,k)$ which depends smoothly on all its arguments.
\\

Due to the smoothness requirement imposed on the system and the coefficients $\mathbb{H}_{ij}$, the system is strongly hyperbolic if and only if the principal symbol, with coefficients given by $\mathbb{A}_{ij}=\mathbb{A}_{ij}^b k_b$, has real eigenvalues and a complete set of eigenvectors \cite{Reula-Rubio2017}.

\subsection{Birefringence}

The dispersion relation given in (\ref{eq:Polinomio}) generally depends on the polarization of the electromagnetic wave when it propagates in an anisotropic medium. This phenomenon is commonly known as \textit{ birefringence}, and is a typical characteristic of most non-linear theories of electromagnetism. 

The standard program to compute how birefringence arises in a particular electromagnetic theory assumes a uniform, constant background electromagnetic field $\Bar{F}_{\mu\nu}$ on a flat background in four spacetime dimensions, and studying plane-wave perturbations $f_{\mu\nu}$ of it with very small amplitudes. The equations in the weak field limit are determined from a perturbed lagrangian which depends on $f_{\mu\nu}$ and derivatives with respect to the electromagnetic invariants. Setting a complex plane-wave Ansatz of the form 
\[
    a_\mu = \epsilon_\mu(k)e^{ik^\nu x_\nu}
\]
with $f_{\mu\nu} := 2\partial_{[\mu}a_{\nu]}$ it is possible to get a dispersion relation by asking the determinant of the set of perturbed equations to vanish. Such a condition can generically be expressed as \citep{Bialynicki-Birula:1984daz}
\begin{equation}
    k^2 \propto \lambda_\pm,
\end{equation}
where
\begin{equation}\label{biref-indeces}
    \lambda_\pm = \frac{1}{2}\frac{2F\Gamma - (\chi_{FF}+\chi_{GG})\pm \sqrt{\delta}}{G^2\Gamma + (F\chi_{GG}-G\chi_{FG})-1},
\end{equation}
\begin{equation}\label{chis}
    \chi_{FF} = \frac{\mathcal{L}_{FF}}{\mathcal{L}_{F}}, \quad
    \chi_{GG} = \frac{\mathcal{L}_{GG}}{\mathcal{L}_{F}}, \quad
    \chi_{FG} = \frac{\mathcal{L}_{FG}}{\mathcal{L}_{F}}\,,
\end{equation}
and
\begin{eqnarray*}
    \Gamma &=& \chi_{FF}\chi_{GG}-\chi_{FG}^2 \,,\\
    \delta &=& (\chi_{FF}-\chi_{GG}-2F\Gamma)^2 + (2\chi_{FG}-2G\Gamma)^2.
\end{eqnarray*}
The functions $\lambda_\pm$ given in (\ref{biref-indeces}) are called \textit{birefringence indices}, and they give an operative way to quantify the strength of birefringence in a given theory. In particular, $\lambda_\pm \equiv 0$ for theories not exhibiting birefringence, and the consequent dispersion relation is the one for the standard wave equation; i.e., $k^\mu k_\mu=0$. One example of this is Maxwell Electromagnetism, in which perturbations propagate at the speed of light, following null light cones precisely given by such dispersion relation. For the case of nonlinear theories, instead, as photon propagation depends on the electromagnetic fields, different polarization states might propagate following quite different light cones, consequently changing the effective dispersion relation, thus giving rise to nonzero birefringence indices. Nevertheless, not every nonlinear theory shows birefringence. In fact, it was shown by Boillat in \citep{Boillat:1966eyw,Boillat:1970gw} that Born-Infeld electrodynamics does not exhibit birefringence, this being also the case of the linear classical Maxwell theory. Moreover, Boillat and Plebanski showed that this theory is the only nonlinear extension of Electromagnetism which admits a weak-field limit for which there is no birefringence. Of course, one may construct further examples of nonlinear theories without such a limit, but they will not remain duality invariant. This interesting effect has been inspected before for very particular scenarios \cite{Bir1,Bir2}, like in static black hole configurations. In this work we go beyond, and numerically study, for the first time, how birefringence is dynamically manifested from numerical simulations, and compare it with other nonlinear extensions.

\section{Proof of well-posedness}
\label{sec-Wellposedness}

In this section, we show that ModMax Electrodynamics admits a well-posed initial-value formulation. To do so, we make use of a geometric criterion for general nonlinear electromagnetic theories, developed in \citep{Reula-Abalos20}. After studying in detail the characteristic structure of the theory, we prove the validity of an inequality as a necessary and sufficient condition for ensuring symmetric hyperbolicity and, therefore, well-posedness of its formulation.

\subsection{Characteristic structure and birefringence}

In order to study the characteristic structure of the theory (and thus how information propagates), we inspect the evolution of plane-wave modes from the corresponding dispersion relation. For the particular case of NLED with lagrangian density $\mathcal{L}(F,G)$, we impose plane-wave solutions with wave covector $k_\mu$ as an Ansatz to Eqs. (\ref{eq:EcENL})-(\ref{eq:EcENL2}). It was proven in \cite{Reula-Abalos20} that the set of covectors $k_\mu$ satisfy the dispersion relation 
\begin{equation}
\label{eq:Polinomio}
    a k^4 + Q \ell^2 k^2 + R \ell^4 = 0,
\end{equation}
where $k^2\equiv g^{\mu \nu} k_\mu k_\nu$, $\ell^2\equiv {F^{\mu}}_\nu F^{\lambda \nu} k_\mu k_\lambda$ and
\begin{eqnarray}
\label{eq:PrincipioConos}
    a &=& 1+8\chi_{FG} G - 8\chi_{GG} F - R G^2 , \nonumber\\
    Q &=& 8 (\chi_{FF} + \chi_{GG} - R F/2) ,\nonumber \\
    R &=& 4(\chi_{FF} \chi_{GG} - \chi_{FG}^2) , \nonumber
\end{eqnarray}
while the functions $\chi_{XY}$ have been introduced in equation (\ref{chis}).
Moreover, the polynomial given in (\ref{eq:Polinomio}) can be factorized, getting
\begin{equation}
\label{eq:PolinomiosEfectivos}
      (g_1^{\mu \nu} k_\mu k_\nu)(g_2^{\mu \nu} k_\mu k_\nu) = 0\,.
\end{equation}
The tensors $g_1^{\mu \nu}$ and $g_2^{\mu \nu}$ are both Lorentzian, and they act as \textit{effective metrics}, as both conditions coincide with the dispersion relations of the wave equation in ``fictitious'' spacetimes with metrics $g^1_{\mu\nu}$ and $g^2_{\mu\nu}$, respectively. These metrics can be explicitly computed, yielding
\begin{eqnarray}
    \label{eq:MetricaEfectiva1}
    g_1^{\mu \nu} &=& a g^{\mu \nu} + b_1 {F^\mu}_\lambda F^{\nu \lambda} ,\nonumber \\
    \label{eq:MetricaEfectiva2}
    g_2^{\mu \nu} &=& g^{\mu \nu} + \frac{b_2}{a} {F^\mu}_\lambda F^{\nu \lambda} ,\nonumber
\end{eqnarray}
where
\begin{eqnarray*}
    b_1 &=& \frac{Q+\sqrt{\Delta}}{2}\,, \\
    b_2 &=& \frac{Q-\sqrt{\Delta}}{2}\,, \\
    \Delta &=& Q^2 - 4 a R.
\end{eqnarray*}
It was proven in \cite{Boillat1970} that the roots of (\ref{eq:PolinomiosEfectivos}) always exist, as the discriminant is a sum of squares, namely, $\Delta = 4 \left(N_1^2 + N_2^2 \right)$, with
\begin{eqnarray}
\label{eq:FinalConos}
      N_1 &=& \chi_{FF} - \chi_{GG} - R F\,,\nonumber\\
      N_2 &=& 2\chi_{FG} - RG. 
\end{eqnarray}

The study of the propagation cones associated with the background metric and both effective metrics introduced before is crucial for the hyperbolicity of the system, as it can be elucidated in what follows.

\subsection{Symmetric--hyperbolicity}

The following theorem (proved in \cite{Reula-Abalos20}) relates the geometric structure of NLED theories and their symmetric hyperbolicity:
\\
\\
$\textbf{Theorem I.}$ \textit{An electromagnetic theory with lagrangian density $\mathcal{L}(F,G)$ is symmetric-hyperbolic if and only if the light cones of the corresponding effective metrics intersect. Moreover, this condition holds if and only if} 
\begin{equation}
\label{eq:TeoSimHip}
1+2\left((FN_1+GN_2)-\sqrt{N_1^2+N_2^2}\sqrt{F^2+G^2}\right)>0, 
\end{equation}
\textit{where $N_1$ y $N_2$ are defined in (\ref{eq:FinalConos})}.
\\
\\
By means of the above criterion, we get the following result:
\\
\\
$\textbf{Theorem II.}$ \textit{ModMax Electrodynamics is symmetric-hyperbolic, and therefore it admits a well-posed initial-value problem.}
\\
\\
$\textbf{Proof}$. We prove this by direct computation of the left-hand side of (\ref{eq:TeoSimHip}). From the lagrangian density (\ref{eq:LagrangianoModMax}), the invariants $\xi_i$ are
\begin{eqnarray}
\chi_{FF}&=&\frac{\sinh(\gamma)\left(\frac{1}{\sqrt{F^2+G^2}}-\frac{F^2}{(F^2+G^2)^{\frac{3}{2}}}\right)}{2\left(-\cosh(\gamma)+\sinh(\gamma)\frac{F}{\sqrt{F^2+G^2}} \right)}, \nonumber \\
\chi_{FG}&=&\frac{-\sinh(\gamma)\frac{FG}{(F^2+G^2)^{\frac{3}{2}}} }{2\left(-\cosh(\gamma)+\sinh(\gamma)\frac{F}{\sqrt{F^2+G^2}} \right)}, \nonumber \\
\chi_{GG}&=&\frac{\sinh(\gamma)\left(\frac{1}{\sqrt{F^2+G^2}}-\frac{G^2}{(F^2+G^2)^{\frac{3}{2}}}\right)}{2\left(-\cosh(\gamma)+\sinh(\gamma)\frac{F}{\sqrt{F^2+G^2}} \right)}, \nonumber
\end{eqnarray}
which implies that $R\equiv0$. Then, the functions $N_i$ are explicitly given by
\begin{equation}
N_1=\frac{\alpha_1}{\mathcal{L}_F}, \nonumber\qquad N_2=\frac{\alpha_2}{\mathcal{L}_F},\nonumber
\end{equation}
with
\begin{eqnarray}
\alpha_1&=&\frac{1}{2}\sinh(\gamma)\left(\frac{G^2-F^2}{(F^2+G^2)^{\frac{3}{2}}}\right), \nonumber \\
\alpha_2&=&-\sinh(\gamma)\left(\frac{FG}{(F^2+G^2)^{\frac{3}{2}}}\right). \nonumber
\end{eqnarray}

In ModMax, $\mathcal{L}_F<0$ and $\left|F/\sqrt{F^2+G^2}\right| \leq 1$, and a straightforward calculation shows that the condition (\ref{eq:TeoSimHip}) is equivalent to 
\begin{equation}\label{last-sh-condition}
\frac{\sinh(\gamma)-\sinh(\gamma)\frac{F}{\sqrt{F^2+G^2}}}{\cosh(\gamma)-\sinh(\gamma)\frac{F}{\sqrt{F^2+G^2}}} < 1\,.\nonumber
\end{equation}
The above condition is always true, as
\begin{eqnarray}
    \sinh(\gamma)&<&\cosh(\gamma), \quad \gamma\in\mathbb{R}, \quad \mbox{and} \nonumber\\
    F &<& \sqrt{F^2+G^2}.\nonumber
\end{eqnarray}
Then, inequality (\ref{eq:TeoSimHip}) holds in ModMax. $\hfill\Box$

Despite its simplicity, Theorem II constitutes a important result, which motivates a deeper exploration of the dynamical properties of ModMax, even in the highly nonlinear regime. It is notable that ModMax remains symmetric-hyperbolic for \textit{any} value of $\gamma$, unlike the majority of nonlinear extensions which, if they are, they are symmetric-hyperbolic only in a region of their parameter space. This condition sorts ModMax proposal out from the rest of extensions. Also, it is important to recall at this point that the condition of symmetric-hyperbolicity we have shown for ModMax is stronger than just well-posedness (or strong-hyperbolicity), and this result allows for long-time stable numerical simulations. 

In what follows, we explore the validity of a series of inequalities relating the energy, charge and angular momentum of the theory. 

\section{Bekenstein bounds}
\label{sec-BekensteinBounds}

Complementary to the hyperbolicity analysis displayed in the previous section, here we study some features of the nonlinear dynamics in ModMax. In particular, we wonder if there exist any relationship between the evolution of physical quantities characterizing the theory, like electromagnetic energy, electric charge and angular momentum. As it was previously motivated, we will elucidate this by assessing the validity of the universal Bekenstein bounds. 

The first inequality studied in this section, given by Eq. (\ref{eq:DesigualdadSinJ}), relates the electric charge and the electromagnetic energy in a bounded region of space, and was previously proven to hold for Maxwell and Born-Infeld theories. Here we point out that this version of the inequality is not well formulated for general NLED theories, prompting us to ``rephrase'' it in a more consistent way; finally proving its validity for spherically symmetric configurations. After that, we analyze the relation between angular momentum, energy and size, and show that the relation (\ref{eq:DesigualdadSinQ}) is valid in ModMax for a generic configuration of the electromagnetic fields. As a straightforward corollary, we show the validity of the general inequality (\ref{eq:DesigualdadSinS}) in spherical symmetry.

\subsection{Inequality between charge, energy and size}

Let $\Omega\subset\mathbb{R}^3$ be a bounded region of space with electric charge $Q$. Let $\mathcal{R}$ be the radius of the smallest sphere $B_{\mathcal{R}}$ that contains $\Omega$. This is one possible measure of the size of $\Omega$\footnote{A discussion on other measures of size for bounded domains can be found in \cite{DainPRL14}.}. With this, it is conjectured that the inequality 
\begin{equation}\label{des-QE-vieja}
\frac{Q^2}{8 \pi \mathcal{R}} \leq \mathcal{E},
\end{equation}
holds in $\Omega$. The motivation of the above inequality is the following. Any region of space storing an amount of electric charge $Q$ must contain a minimum energy, which is proportional to $Q^2$ and inversely proportional to its size. The validity of this inequality was shown by Dain for Maxwell's Electromagnetism \cite{Dain2015}, and explored two years later by Peñafiel et. al. for Born-Infeld theory \cite{Penafiel2017}. Shortly after, it was pointed out that this version of the inequality is not generally true for nonlinear theories, and a ``generalised'' one was proposed in \citep{PhysRevD.98.049902}, which reads
\begin{equation}
\label{eq:DesigualdadCargaENL}
\mathcal{E} \geq \frac{Q}{2} \phi(\mathcal{R}).
\end{equation}
Here, $\phi(\mathcal{R})$ is the electrostatic potential in $\Omega$; i.e., the work needed to bring a unit of charge from infinity to $\Omega$. In fact, inequality (\ref{eq:DesigualdadCargaENL}) was proved for Born-Infeld in spherical symmetry \cite{PenafielErratum2018}. In order to explore the relation between energy, charge and size in ModMax, we first analyzed the original version of the inequality, given by Eq. (\ref{des-QE-vieja}), and showed that it does it is not necessarily true in ModMax. A counterexample of it is displayed in Appendix \ref{app-counterex}. 

Let us now inspect inequality (\ref{eq:DesigualdadCargaENL}) in spherical symmetry. The total electromagnetic energy is always greater than the electromagnetic energy of the corresponding electrostatic configuration, for which we can restrict our study, without loss of generality, to purely electrostatic configurations, i.e., $\mathbf{B}=\mathbf{0}$ and $\mathbf{E}=E(r)\mathbf{\hat{r}}$, because of the spherical symmetry. The total energy of the system is
\begin{equation}
\mathcal{E}= \int_{\mathbb{R}^3}^{} u ,\nonumber
\end{equation}
where $u$ is the energy density in ModMax, which for electrostatic configurations yields
\begin{equation}
u=\frac{e^{\gamma}}{2}\textbf{E}^2.\nonumber
\end{equation}
We split the total energy as a sum of two terms: the energy $\mathcal{E}(r<\mathcal{R})$ inside $B_\mathcal{R}$, plus the energy $\mathcal{E}(r>\mathcal{R})$ outside of it, where $r$ is a radial coordinate measured from the center of $B_\mathcal{R}$. In particular, since $u\geq 0$, it is $\mathcal{E} \geq \mathcal{E}(r>\mathcal{R})$. On the other hand, the electrostatic potential is given by
\begin{equation}
\label{eq:Potencial}
\phi(\mathcal{R})= \int_{\mathcal{C}}^{} \mathbf{E} \,\cdot \mathbf{dl},\nonumber
\end{equation}
where $\mathcal{C}$ is any path from infinity to $r = \mathcal{R}$, which in spherical symmetry reads
\begin{equation}
\phi(\mathcal{R})= \int_{\mathcal{R}}^{\infty} E(r) \,dr.\nonumber
\end{equation}
From Gauss's theorem, which is also valid in NLED, we get 
\begin{equation}
\int_{\partial B_\mathcal{R}}^{} \mathbf{D} \cdot \mathbf{\hat{n}} \,dS= \int_{B_\mathcal{R}}^{} \rho,\nonumber
\end{equation}
where the integral on the left is over the boundary of $B_\mathcal{R}$ with outward normal $\mathbf{\hat{n}}$. Because of the symmetry, we have that $\mathbf{D}=D(r)\mathbf{\hat{r}}$, which directly implies that $\mathbf{D}=e^{\gamma}\mathbf{E}$. Solving for the electric field, we have 
\begin{equation}
\mathbf{E}= \frac{e^{-\gamma} Q }{4\pi r^2} \mathbf{\hat{r}},\nonumber
\end{equation}
and from Eq. (\ref{eq:Potencial}), we finally get
\begin{eqnarray}
\label{eq:EFuerayPhi}
\mathcal{E}(r>\mathcal{R})&=&\frac{e^{-\gamma} Q^2}{8\pi \mathcal{R}},\nonumber\\
\phi(\mathcal{R})&=&\frac{e^{-\gamma} Q }{4\pi \mathcal{R}}\;.\nonumber
\end{eqnarray}
Thus, 
\begin{equation}
    \mathcal{E}(r>\mathcal{R})= \frac{1}{2}Q \phi(\mathcal{R}),
\end{equation}
and since $\mathcal{E} \geq \mathcal{E}(r>\mathcal{R})$, the inequality (\ref{eq:DesigualdadCargaENL}) holds.

\subsection{Inequality between angular momentum, energy and size}

We now proceed with the analysis of inequality (\ref{eq:DesigualdadSinQ}) in ModMax. This inequality relates the angular momentum, total energy and size of an bounded electromagnetic configuration. It was shown to hold in Maxwell's Electromagnetism \cite{Dain2015}, and also in Born-Infeld theory \cite{Penafiel2017}. Such inequality tells us that the total energy of the system is always bounded from below by the angular kinetic energy. Furthermore, as the quotient of both terms is proportional to the mean angular velocity of the system, inequality (\ref{eq:DesigualdadSinQ}) establishes that this quantity must be lower than one (i.e., the speed of light in geometric units). The relation between this inequality and the causality of theories has been studied earlier \cite{Penafiel2017}, concluding that there are nonlinear theories that could satisfy inequality (\ref{eq:DesigualdadSinQ}) but violate causality. 

Here, we show the following theorem.
\\
\\
$\textbf{Theorem IV.}$ \textit{Let $\Sigma\subset\mathbb{R}^3$ be a bounded region in space, and $\mathcal{R}(\Sigma)$ the radius of the smallest sphere that contains $\Sigma$. Then, the inequality
\begin{equation}
\centering
\label{eq:DesigualdadSinQ2}
\mathcal{E}(\Sigma)\mathcal{R}(\Sigma) \geq \lvert \mathcal{J}(\Sigma) \rvert
\end{equation}
holds in ModMax electrodynamics, where $\mathcal{E}(\Sigma)$ and $\mathcal{J}(\Sigma)$ are, respectively, the energy and angular momentum in $\Sigma$.}
\\
\\
$\textbf{Proof.}$ From Eq.~ (\ref{eq:MomentoAngularENL}), the angular momentum in ModMax reads
\begin{equation}
\centering
\label{eq:MomentoAngularModMax}
\mathcal{J}(\Sigma)=\int_{\Sigma}^{} \left(\cosh(\gamma)-\sinh(\gamma) \frac{F}{\sqrt{\eta}} \right) (\textbf{x} \times (\textbf{E} \times \textbf{B}))\cdot \hat{k} \nonumber 
\end{equation}
where $\eta\equiv F^2+G^2$. The right-hand side of the above formula can be bounded by means of the Cauchy–Schwarz inequality. In fact, we have that
\begin{equation}
    \lvert (\textbf{x} \times (\textbf{E} \times \textbf{B}))\cdot \hat{k} \rvert \leq \lvert \textbf{x}\rvert \lvert \textbf{E}\rvert \lvert \textbf{B}\rvert \lvert \hat{k} \rvert = \lvert \textbf{x}\rvert \lvert \textbf{E}\rvert \lvert \textbf{B}\rvert,   \nonumber
\end{equation}
and thus, 
\begin{equation}
\label{eq:CotaModuloJ}
\lvert \mathcal{J}(\Sigma) \rvert \leq \int_{\Sigma}^{} \left|\cosh(\gamma)-\sinh(\gamma)\frac{\textbf{B}^2-\textbf{E}^2}{\sqrt{\eta}} \right| \lvert \textbf{x}\rvert \lvert \textbf{E}\rvert \lvert \textbf{B}\rvert.\nonumber
\end{equation}
With this, we can bound from below the difference between the left and right hand sides of inequality (\ref{eq:DesigualdadSinQ2}) using that the electromagnetic energy density in ModMax is
\begin{equation}
u=\frac{1}{2}(\textbf{E}^2+\textbf{B}^2)\left(\cosh(\gamma)-\sinh(\gamma)\frac{\textbf{B}^2-\textbf{E}^2}{\sqrt{\eta}}\right).\nonumber
\end{equation}
We get
\begin{eqnarray}
\mathcal{E}(\Sigma) &-& \frac{\lvert \mathcal{J}(\Sigma)\rvert}{\mathcal{R}} \nonumber\\
& \geq & \int_{\Sigma}
\frac{\textbf{E}^2 + \textbf{B}^2}{2} \left(\cosh(\gamma)-\sinh(\gamma)\frac{\textbf{B}^2-\textbf{E}^2}{\sqrt{\eta}}\right) \, \nonumber\\
&-&\int_{\Sigma} \frac{\lvert\textbf{x}\rvert}{\mathcal{R}}\lvert\textbf{E}\rvert \lvert\textbf{B}\rvert \left( \cosh(\gamma)-\sinh(\gamma)\frac{\textbf{B}^2-\textbf{E}^2}{\sqrt{\eta}}\right)\,  \nonumber\\
&\geq& \int_{\Sigma}^{} \frac{1}{2} \left(\cosh(\gamma)-\sinh(\gamma)\frac{\textbf{B}^2-\textbf{E}^2}{\sqrt{\eta}}\right)\nonumber\\
&\times& \left( \textbf{E}^2+\textbf{B}^2-2\frac{\lvert\textbf{x}\rvert}{\mathcal{R}}\lvert\textbf{E}\rvert \lvert\textbf{B}\rvert \right) \, \nonumber
\end{eqnarray}
Now, since \textbf{x} $\in\Sigma$, it is $\lvert \textbf{x} \rvert / \mathcal{R} \leq 1$, which implies that
\begin{eqnarray}\label{eq:ineq-fin}
\mathcal{E}(\Sigma) &-& \frac{\lvert \mathcal{J}(\Sigma)\rvert}{\mathcal{R}} \\
&\geq& \int_{\Sigma}^{} \left(\cosh(\gamma)-\sinh(\gamma)\frac{\textbf{B}^2-\textbf{E}^2}{\sqrt{\eta}}\right)\frac{\left(\textbf{E}^2-\textbf{B}^2 \right) ^2}{2} \nonumber.
\end{eqnarray}
Since $\gamma\geq 0$ and
\begin{equation}
\left|\frac{\textbf{B}^2-\textbf{E}^2}{\sqrt{\eta}}\right| \leq 1, \nonumber
\end{equation}
the integral on the right-hand side of (\ref{eq:ineq-fin}) results non negative, finally getting the inequality (\ref{eq:DesigualdadSinQ2}). $\hfill\Box$

Let us observe that for proving Theorem IV, there is no need to assume that the region $\Sigma$ is free of charge; i.e., this inequality is independent of the amount of electric charge there stored. Nevertheless, if $\Sigma$ is charged, a better estimate for the energy can be given, as we will see in what follows. 

\subsection{Inequality between charge, energy and angular momentum}

As a simple corollary of the two previous inequalities, we now study the general case given by Dain's inequality (\ref{eq:DesigualdadSinS}). In fact, let us assume ModMax electrodynamics, sourced by a charge density $\rho$ with compact support. In analogy with the first inequality shown in the previous section, here we wonder the validity of the bound
\begin{equation}
\label{eq:DesigualdadGeneralENL}
\mathcal{E} \geq \frac{Q}{2}\phi(\mathcal{R}) + \frac{\lvert \mathcal{J}(\Sigma) \rvert}{\mathcal{R}}\,,
\end{equation}
in spherical symmetry. As it was done before, the total energy $\mathcal{E}$ of the system can be expressed as 
\begin{equation}
\mathcal{E} = \int_{\mathbb{R}^3/\Sigma}^{} u \,+ \int_{\Sigma}^{} u \,.\nonumber
\end{equation}
The second integral in the above formula is exactly $\mathcal{E}(\Sigma)$, which can be bounded from below by $\vert \mathcal{J}(\Sigma) \vert / \mathcal{R}$, by virtue of inequality (\ref{eq:DesigualdadSinQ2}), yielding
\begin{eqnarray}
\mathcal{E} &\geq& \int_{\mathbb{R}^3/\Sigma}^{} u \, + \frac{\vert \mathcal{J}(\Sigma) \vert}{\mathcal{R}} \nonumber \\
&=& \mathcal{E}(r>\mathcal{R}) + \frac{\vert \mathcal{J}(\Sigma) \vert}{\mathcal{R}},\nonumber
\end{eqnarray}
where in the second line we have use the spherical symmetry. But for this particular case, it was proven that $\mathcal{E}(r>\mathcal{R}) = (Q/2) \phi (\mathcal{R})$, so that 
\begin{equation}
\mathcal{E} \geq \frac{Q}{2}\phi(\mathcal{R}) + \frac{\lvert \mathcal{J}(\Sigma) \rvert}{\mathcal{R}},\nonumber
\end{equation}
or
\begin{equation}
\mathcal{E}^2 \geq \left(\frac{Q}{2}\phi(\mathcal{R})\right)^2 + \left(\frac{\lvert \mathcal{J}(\Sigma) \rvert}{\mathcal{R}}\right)^2 +  \frac{Q}{\mathcal{R}} \phi(\mathcal{R}) \vert \mathcal{J}(\mathcal{R})  \vert.\nonumber
\end{equation}
Since the last term on the right-hand side is non-negative, we finally get
\begin{equation}
\mathcal{E}^2 \geq \left(\frac{Q}{2}\phi(\mathcal{R})\right)^2 + \left(\frac{\lvert \mathcal{J}(\Sigma) \rvert}{\mathcal{R}}\right)^2.\nonumber
\end{equation}

\section{Numerical simulations}
\label{sec-Numerics}

\subsection{Evolution equations in ``slab'' symmetry}

We numerically evolve the ModMax equations over a flat\footnote{Naturally, all the procedure shown for the evolution equations used here can be straightforwardly generalized to curved backgrounds.} background $\mathcal{M}=(\mathbb{R}^4,\eta_{ab})$, where $\eta_{ab}$ is the Minkowski metric. In order to set evolution equations from the covariant system (\ref{eq:EcENL})-(\ref{eq:EcENL2}), we pick a spatial hypersurface $S$ and put local inertial coordinates $\{x^i\}$ on it. Then, we take a time-like vector field $t^a$ such that $t^a v_a = 0$ for any $v_a\in T^*_pS$, and extend the coordinates $\{x^i\}$ so that they are constant along the integral curves of $t^a$. The evolution of the time coordinate will then be driven by a scalar field $t:\mathcal{M}\to\mathbb{R}$, satisfying: (i) $t\equiv 0$ in $S$; (ii) $t^a\nabla_a t = 1$ on $\mathcal{M}$, where $\nabla$ is the connection compatible with $\eta_{ab}$; and (iii) $t^a = (\partial/\partial t)^a$. This construction allows us to rewrite the system of equations for a general NLED theory in the way
\begin{eqnarray}
\label{eq:EcEvolucion}
\partial_t \mathbf{D} &=& \nabla \times \mathbf{H} - \mathbf{J}, \\
\label{eq:EcEvolucion2}
\partial_t \mathbf{B} &=& -\nabla \times \mathbf{E},
\end{eqnarray}
where $\mathbf{J}$ is the electric current density. We see that the equations look the same as Maxwell's equations in continuous media. Certainly, their solutions must also satisfy the constraints
\begin{eqnarray}
\label{eq:EcVinculo}
\nabla \cdot \mathbf{D} &=& \rho, \\
\label{eq:EcVinculo2}
\nabla \cdot \mathbf{B} &=& 0,
\end{eqnarray}
where $\rho$ is the free charge density. As in the case of Maxwell's equations, the sources $(\rho,\,\mathbf{J})$ cannot be arbitrarily given. It is easy to see that, as a consequence of (\ref{eq:EcEvolucion}) and (\ref{eq:EcVinculo}), they must follow the \textit{continuity equation} 
\begin{equation}\label{eq:Ec-Continuidad}
    \frac{\partial\rho}{\partial t} + \nabla\cdot\textbf{J} = 0,
\end{equation}
which stands as an \textit{integrability} condition for the whole system. The nonlinear dependence in the system (\ref{eq:EcEvolucion}) is encoded in the relation between the \textit{dynamical} fields (\textbf{D}, \textbf{B}) and the \textit{fluxes} (\textbf{E}, \textbf{H}). The latter can be expressed as algebraic functions of the former through nonlinear constitutive relations, which can be directly derived from the action. Then, a direct consequence of the hyperbolicity result proven in Theorem II is that the corresponding Cauchy problem is \textit{well-posed} if and only if the initial fields $\mathbf{D}$ and $\mathbf{B}$ are chosen in a way that they satisfy the constraint equations (\ref{eq:EcVinculo})-(\ref{eq:EcVinculo2}), which clearly propagate along evolution. 

For the numerical implementation displayed in this work, we consider the simplest ``slab-symmetric'' case, for which the dynamical fields only depend on one spatial coordinate (which we take to be $z$) and time. Namely, we consider the Ansatz
\begin{equation}\label{eq:Ansatz}
    \mathbf{D} = \mathbf{D}(t,z),\qquad\mathbf{B} = \mathbf{B}(t,z).\nonumber
\end{equation}
As a consequence, the constraint (\ref{eq:EcVinculo2}) implies that $B_z$ cannot depend on $z$; i.e.,
\begin{equation}
    B_z = B_z(t),\nonumber
\end{equation}
while from Eq. (\ref{eq:EcVinculo}) we get 
\begin{equation}\label{eqDz_rho}
    \partial_z D_z = \rho.\nonumber
\end{equation}
A simple choice for the sources is $\bf{J}=0$ and $\rho = \rho(z)$ (with compact support contained in our numerical domain), which trivially satisfy the continuity equation (\ref{eq:Ec-Continuidad}). Furthermore, using the evolution equations (\ref{eq:EcEvolucion})-(\ref{eq:EcEvolucion2}) one more time, we get
\begin{equation}
    B_z = \text{const.},   \nonumber
\end{equation}
and 
\begin{equation}\label{Dz-notime}
    D_z=D_z(z).\nonumber
\end{equation}
Thus, once $\rho$ is given, we can solve for $D_z$ from (\ref{eqDz_rho}) and, without loss of generality, choose $B_z=0$ at $t=0$ (which automatically yields $B_z=0$ for all times).

Here we will numerically verify the validity of inequality (\ref{eq:DesigualdadSinQ}) in \textit{vacuum} ($\bf{J}=0$, $\rho = 0$), for which we can choose $B_z = D_z = 0$. The evolution equations for the remaining fields directly follow from (\ref{eq:EcEvolucion})-(\ref{eq:EcEvolucion2}), reading
\begin{eqnarray}
\label{eq:EcEvolucionComponentes}
\partial_t D_x + \partial_z H_y &=& 0 \nonumber\\
\label{eq:EcEvolucionComponentes2}
\partial_t D_y - \partial_z H_x &=& 0 \nonumber\\
\label{eq:EcEvolucionComponentes3}
\partial_t B_x - \partial_z E_y &=& 0\nonumber\\
\label{eq:EcEvolucionComponentes4}
\partial_t B_y + \partial_z E_x &=& 0\,. 
\end{eqnarray}
This is a nonlinear, homogeneous system of conservation laws of the form
\begin{equation} \label{homogeneus-system}
    \partial_t \mathbf{U} - \partial_z  \mathbf{F}( \mathbf{U}) =  \textbf{0}\,,
\end{equation}
for the variables 
\begin{equation}
 \mathbf{U} = \{D_x,D_y,B_x,B_y\}   
\end{equation}
and fluxes
\begin{equation}
    \mathbf{F}( \mathbf{U}) = \{-H_y,H_x,E_y,-E_x\}.    
\end{equation}

As previously stressed, it is necessary to express the fluxes in terms of the dynamical variables. In this case, such a map is explicit and algebraic, and it can be directly obtained by means of the Hamiltonian formulation. Indeed, from the Hamiltonian density  (\ref{eq:HamiltonianoModMax}) and the constitutive relations (\ref{eq:EHamiltoniano}) we find
\begin{eqnarray}
    H_x &=& \cosh(\gamma) B_x +\sinh(\gamma)\frac{PB_x-2QD_x}{\sqrt{\Delta}}\,,\nonumber \\
    H_y &=& \cosh(\gamma) B_y +\sinh(\gamma)\frac{PB_y-2QD_y}{\sqrt{\Delta}}\,,\nonumber \\
    E_x &=& \cosh(\gamma) D_x -\sinh(\gamma)\frac{PD_x+2QB_x}{\sqrt{\Delta}}\,, \nonumber\\
    E_y &=& \cosh(\gamma) D_y -\sinh(\gamma)\frac{PD_y+2QB_y}{\sqrt{\Delta}}\,,\nonumber
\end{eqnarray}
where $P$, $Q$ and $\Delta$ are given by
\begin{eqnarray*}
    P &:=& \mathbf{D}^2 - \mathbf{B}^2\,, \\
    Q &:=& \mathbf{D}\cdot\mathbf{B}\,, \\
    \Delta &:=& P^2 + 4Q^2.
\end{eqnarray*}
We note at this point that if $\rho\neq 0$, the functions $P$ and $\Delta$ would depend on $D_z$, which has not been defined \textit{a priory} as a dynamical variable. Then, its contribution to the evolution equations should in principle be treated as a source term, and the system would no longer be homogeneous. Alternatively, one could instead consider the ``augmented'' system
\begin{equation}
 \mathbf{U}_a = \{D_x,D_y,D_z,B_x,B_y,B_z\}\,,\nonumber
\end{equation}
with fluxes
\begin{equation}
    \mathbf{F}_a(\mathbf{U}_a) = \{-H_y,H_x,\,0\,,E_y,-E_x,0\},\nonumber
\end{equation}
resulting homogeneous as the one in (\ref{homogeneus-system}). We recall that for the vacuum case, initial data should be set for $D_x$, $D_y$, $B_x$ and $B_y$, keeping the choice $D_z=B_z=0$.

\subsection{Numerical implementation, initial data and boundary conditions}

We evolve the ModMax equations on a cubic domain of edge $L$, whose center coincides with the origin of our Cartesian coordinate system. As was previously pointed out, the corresponding set of evolution equations constitutes a system of conservation laws, and since the corresponding fluxes are nonlinear functions of the dynamical fields, the formation of shocks along evolution is expected \cite{ConsLaw}. Finite-difference schemes are usually not suitable for capturing the formation of discontinuities, much less allow stable propagation of them along evolution. In fact, the fields tend to oscillate with high frequency around the location of the shock, resulting in a poorly resolved simulation. In order to deal with these shortcomings, a wide variety of high-resolution shock-capturing schemes can be implemented for getting an accurate and stable evolution. For our simulations, we implement a fourth-order Runge-Kutta scheme with total variation diminishing, ensuring non-oscillatory shocks and right convergence. We do this by modifying the equations by adding artificial dissipation in the neighborhood of the discontinuity, of the form
\begin{equation}
    \partial_t \mathbf{U} - \partial_z  \mathbf{F}( \mathbf{U}) =  \epsilon\, \partial^2_z\mathbf{U}\,,
\end{equation}
with $\epsilon>0$ small enough. This technique has been extensively explored in the past, mimicking the addition of a ``diffusive-like'' term, and assuring convergence to the right solution through the corresponding entropy conditions \citep{ConsLaw}. We discretized the spatial derivatives using second-order finite difference stencils which satisfy the property of summation by parts \citep{Calabrese:2003vx}. At each time step, we keep track on the energy and the angular momentum, defined in terms of the evolution fields, and use them in order to verify the inequality proven in the previous section.

For the initial configuration, we choose gaussian profiles for the fields, of the form
\begin{equation}
    f(z) = a_0\exp\left[-\frac{(z-z_c)^2}{\sigma^2}\right] + f_0,\nonumber
\end{equation}
where the parameters $(a_0,f_0)$ where set as shown in Table \ref{Tabla}. For all the simulations, we took $L = 40$, $z_c = 0$ and $\sigma = 0.2$.

\begin{table}[h]
\begin{center}
\begin{tabular}{ |p{1cm}||p{1cm}|p{1cm}| }
 \hline
    & $\;\;\;\, a_0$  &$\;\;\;\, f_0$\\
 \hline
 $\;\;\; D_x$ & $\;\;\;\; 1$  & $\;\; 0.01$\\
 \hline
 $\;\;\; D_y$  & $\;\;\;\; 0$  & $\;\;\;\; 0$\\
 \hline
 $\;\;\; B_x$ & $\;\;\;\; 1$  &$\; -0.01$\\
 \hline
 $\;\;\; B_y$  & $\;\;\;\; 0$  &$\;\;\;\; 0$\\
 \hline
\end{tabular}
\end{center}
\caption{Parameters for the initial Gaussian profile of the dynamical fields.}
\label{Tabla}
\end{table}

For simplicity, we impose periodic boundary conditions for all dynamical fields. This implies that in our simulations we have to choose a large enough numerical domain in order to avoid spurious reflections or other boundary effects along evolution. In practice this is not a problem, as nonlinear electrodynamics simulations are quite cheap.

\subsection{Diagnostic quantities}

\subsubsection{Energy and angular momentum}

During the evolution, we keep track of the values of the electromagnetic energy and the angular momentum of the system. Using the symmetries of the Ansatz considered, we can express the energy as
\begin{equation}
\label{eq:NumEnergia}
\mathcal{E}(t)= -\frac{L^2}{2} \int_{-L/2}^{L/2}{(\textbf{E}^2+\textbf{B}^2)\,\mathcal{L}_F} \, dz,
\end{equation}
where
\begin{equation}
\mathcal{L}_F = -\cosh(\gamma)+\sinh(\gamma) \frac{\textbf{B}^2 - \textbf{E}^2}{\sqrt{{\left(\textbf{B}^2 - \textbf{E}^2 \right)}^2+4 \left(\textbf{B} \cdot \textbf{E}\right)^2}}, \nonumber
\end{equation}
and write the components of $\textbf{E}$ in terms of the evolution variables using the Hamiltonian (\ref{eq:HamiltonianoModMax}) and the constitutive relation (\ref{eq:EHamiltoniano}). 

\begin{figure*}
  \centering
  \includegraphics[scale=0.6]{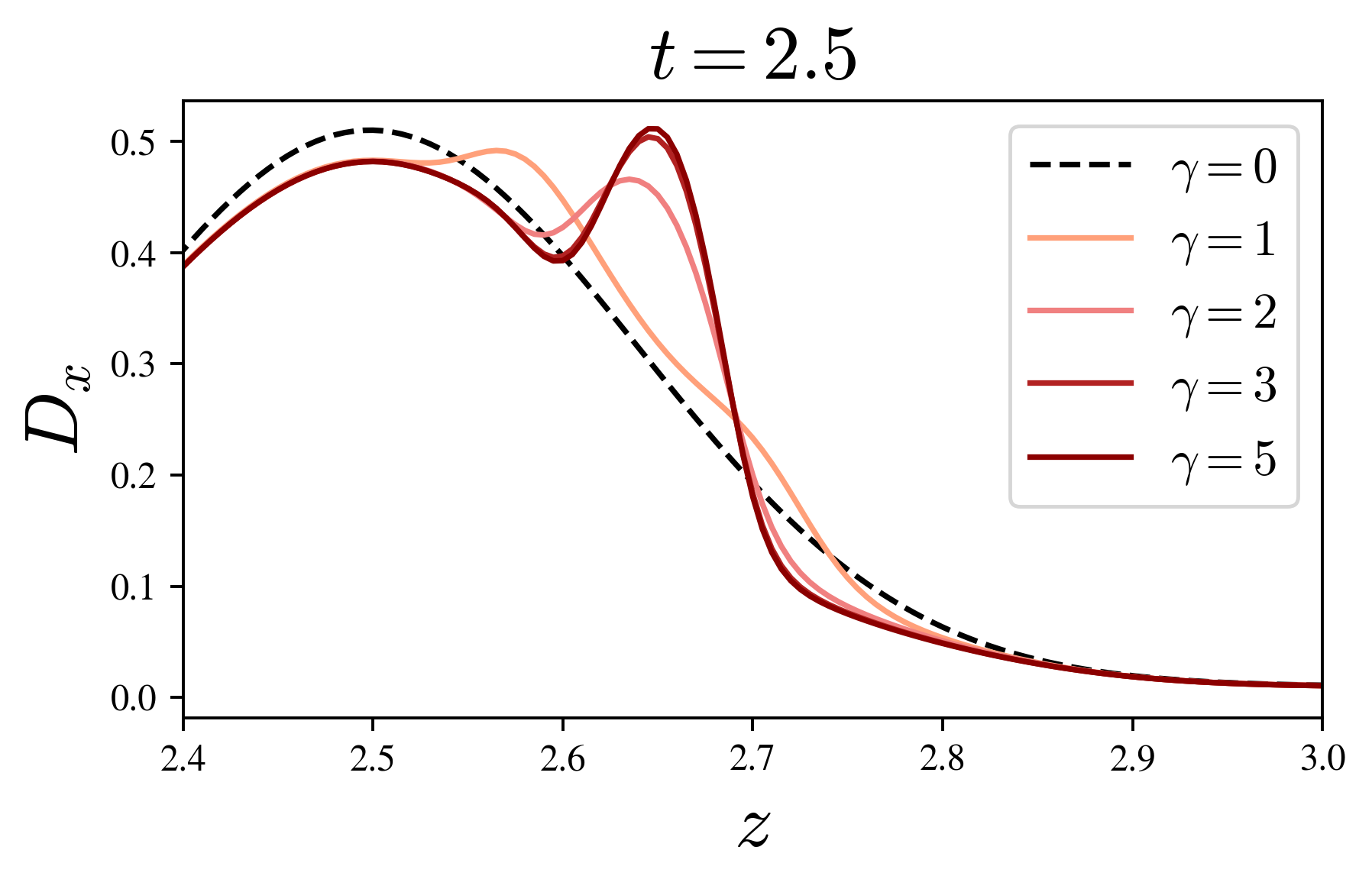}
  \includegraphics[scale=0.6]{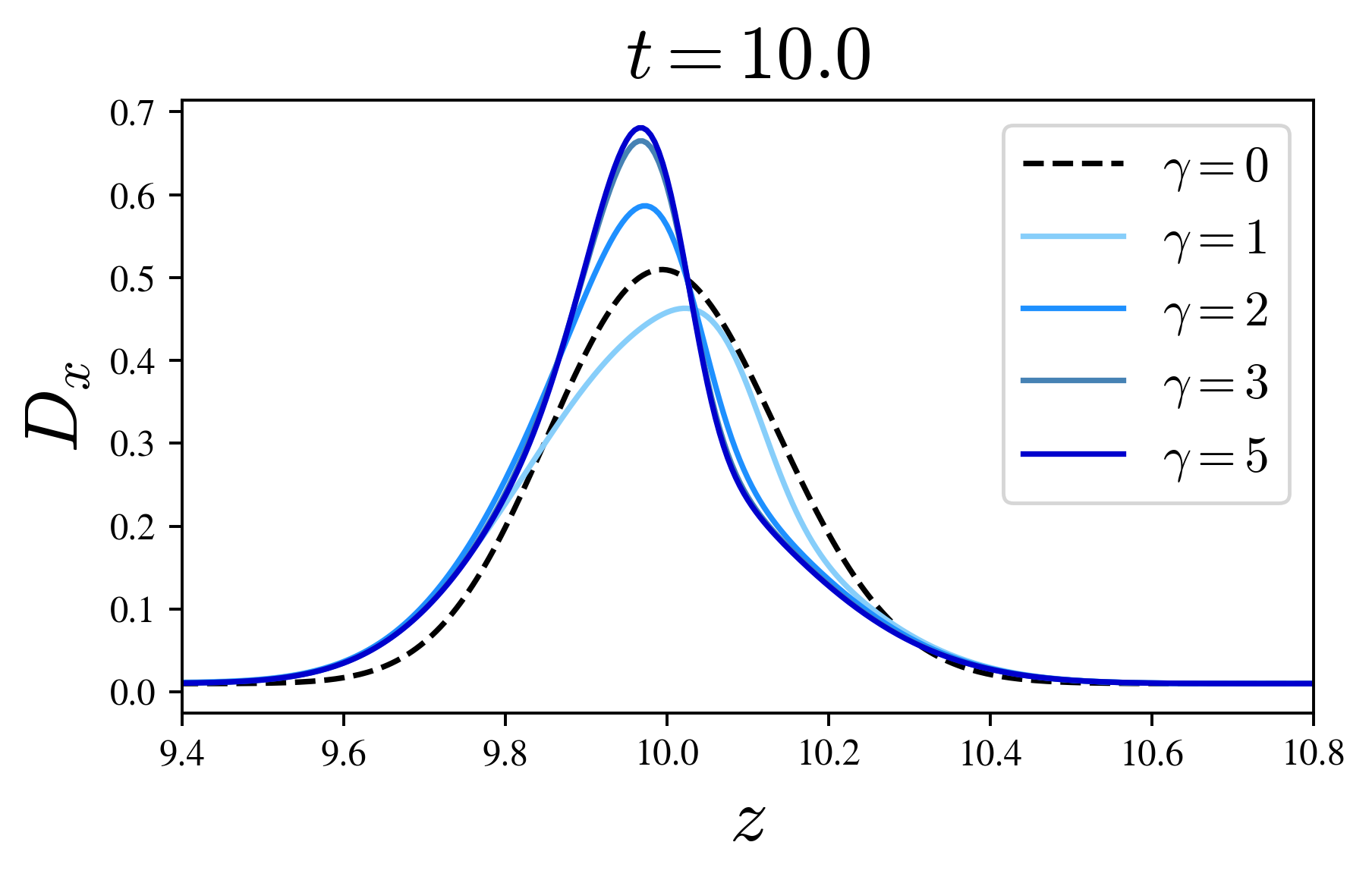}
  \includegraphics[scale=0.6]{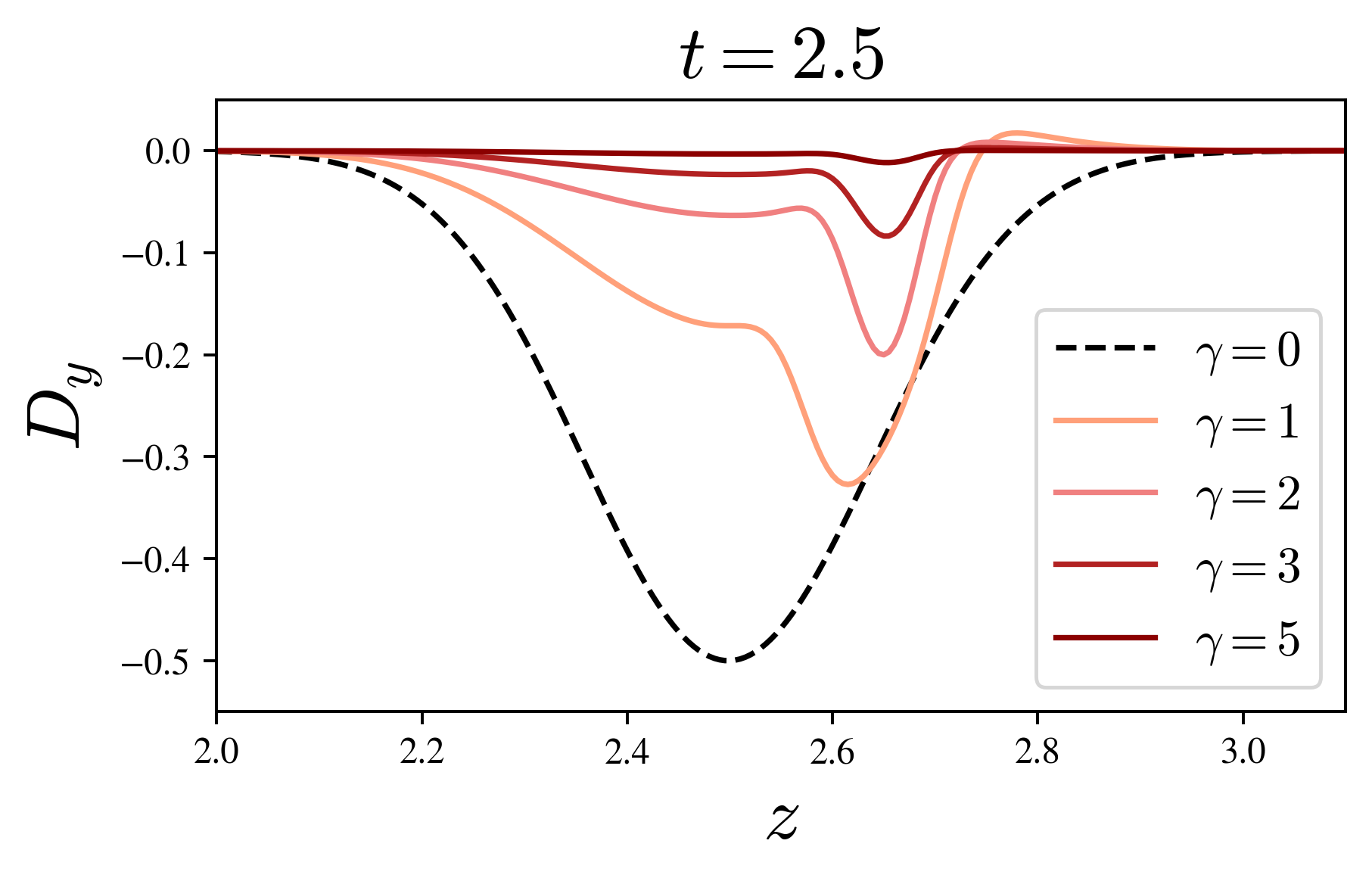}
  \includegraphics[scale=0.6]{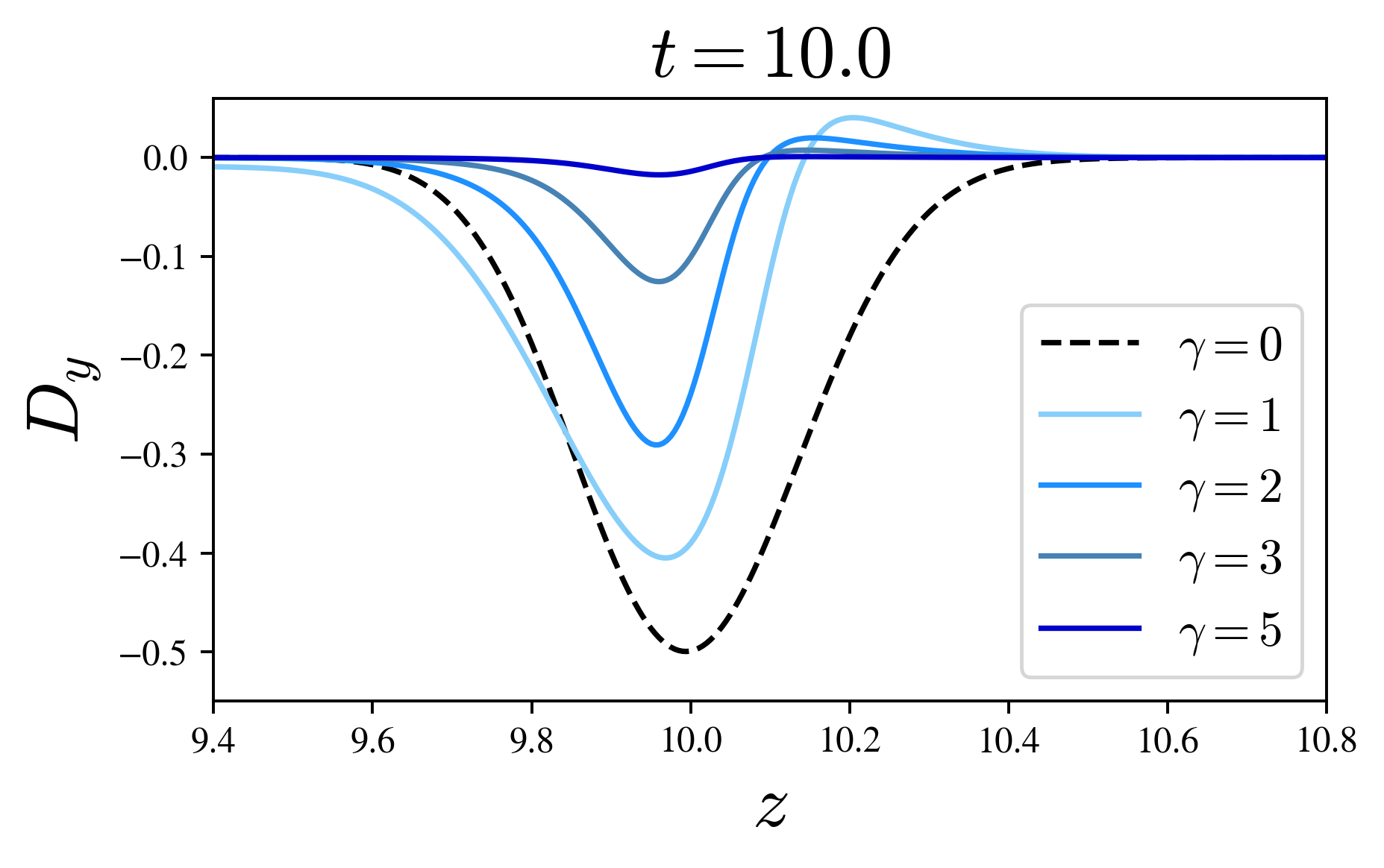}
  \includegraphics[scale=0.6]{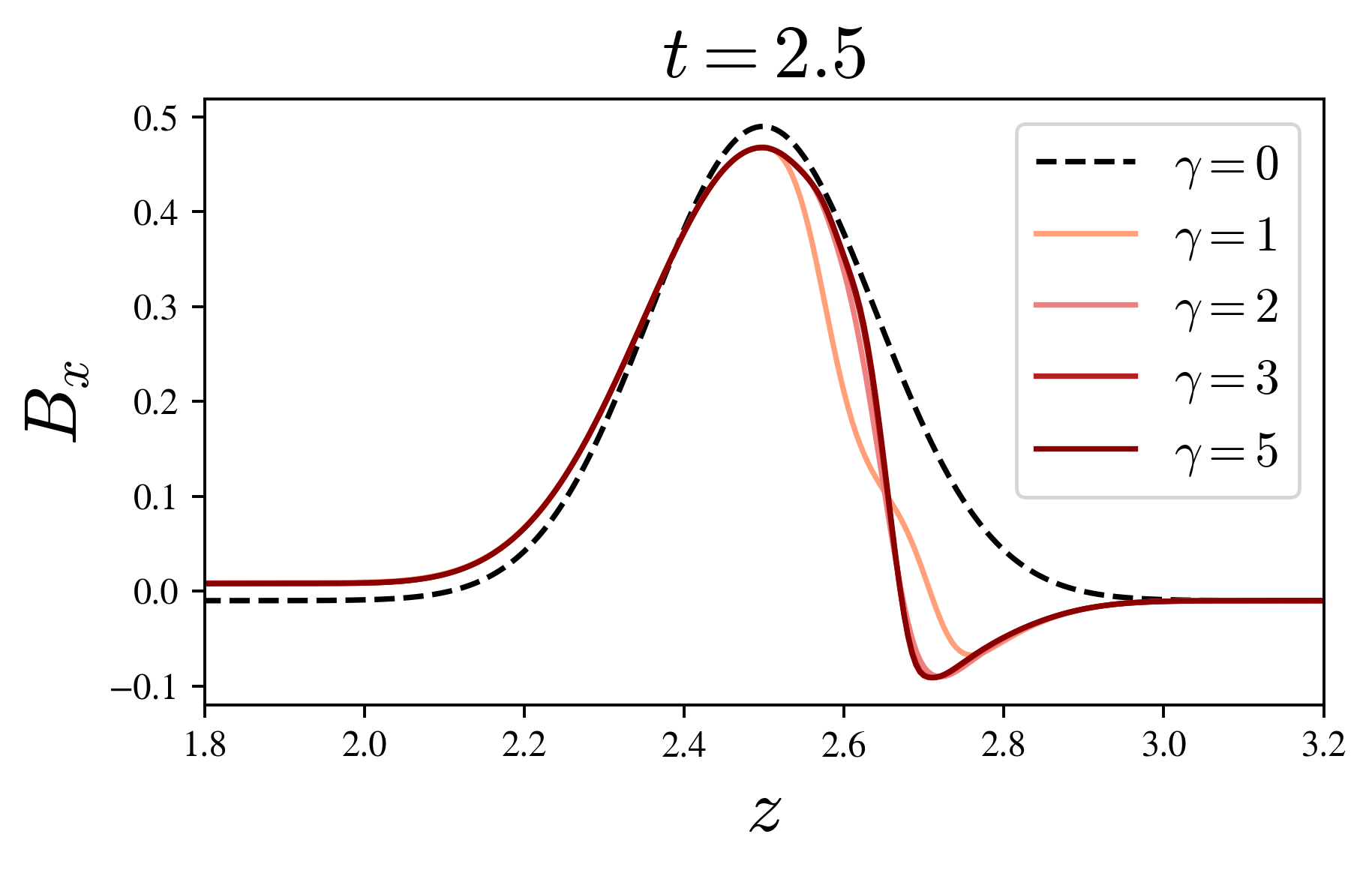}
  \includegraphics[scale=0.6]{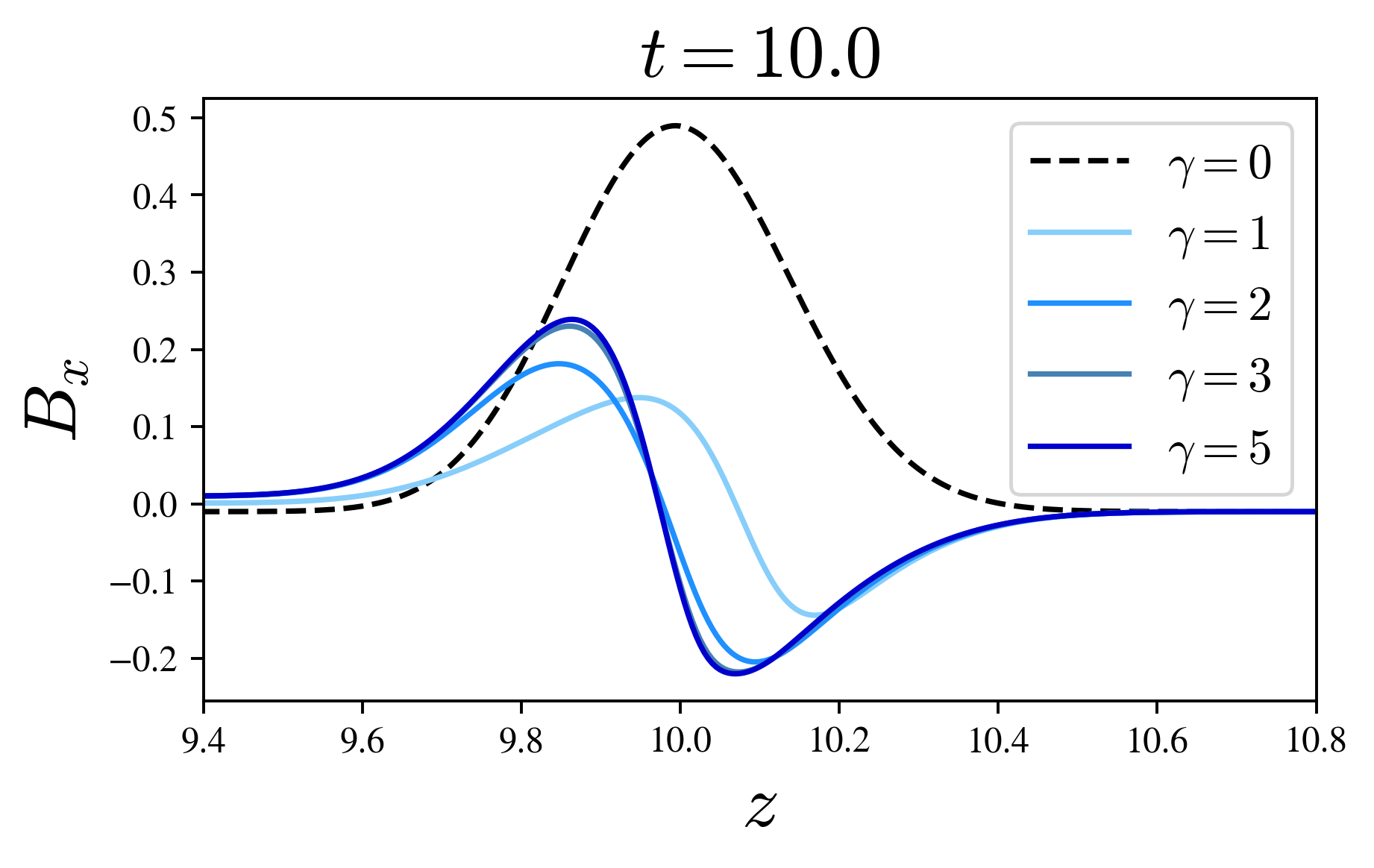}
  \includegraphics[scale=0.6]{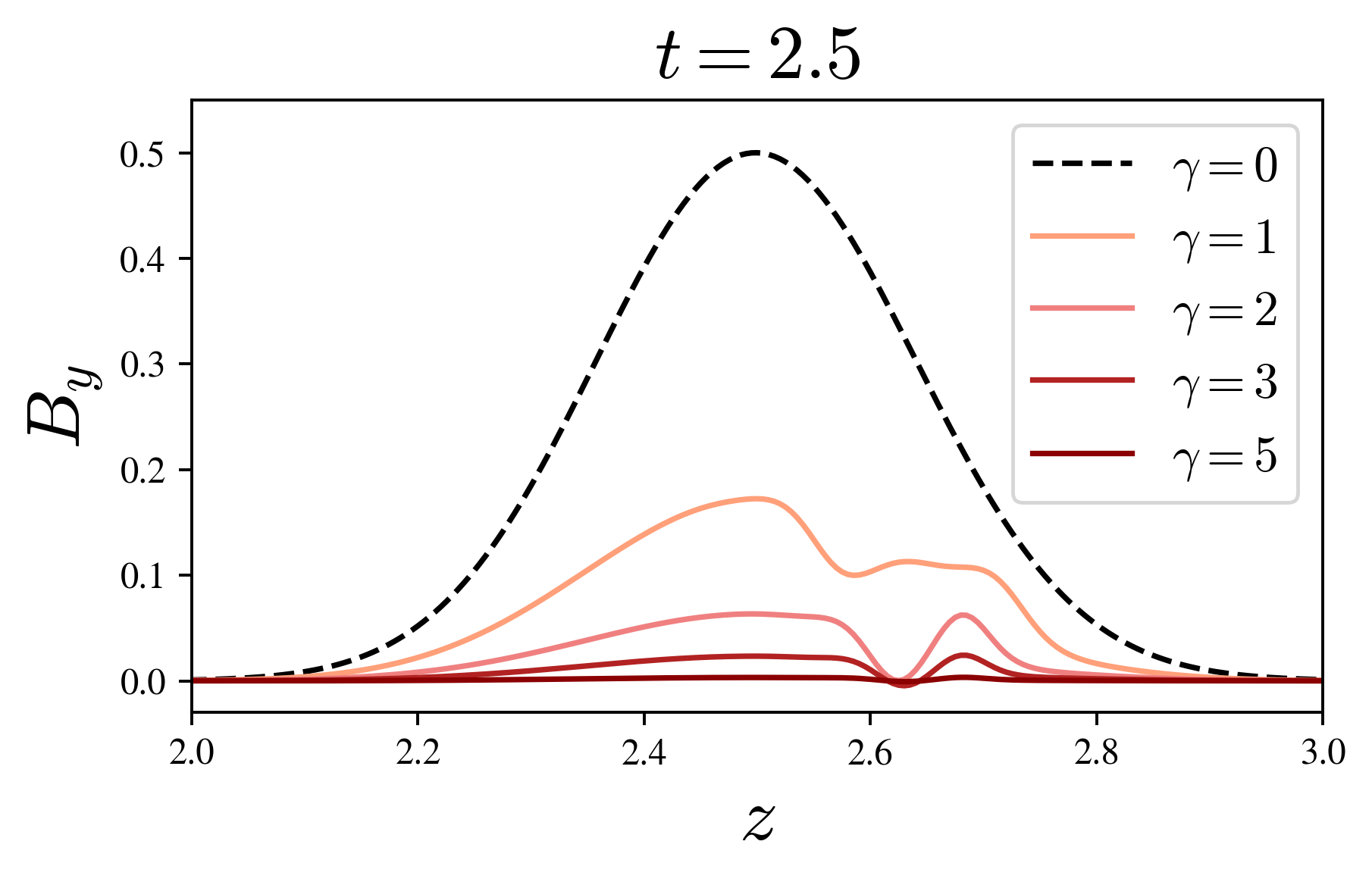}
  \includegraphics[scale=0.6]{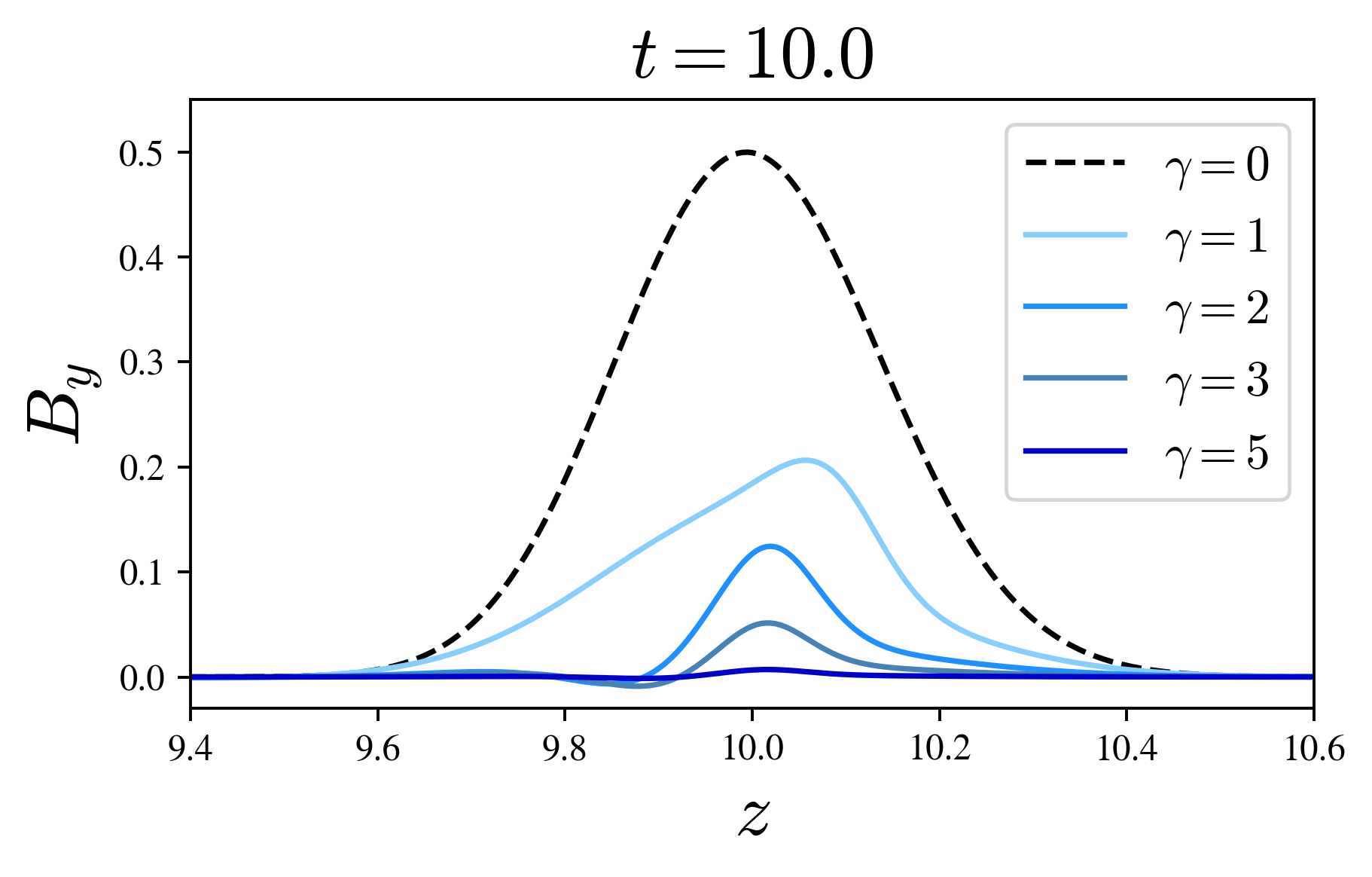}
  \caption{\textit{ModMax evolution.} Snapshots of the evolution of the electromagnetic fields at times $t=2.5$ (left column) and $t=10.0$ (right column), from the initial data specified in Table \ref{Tabla}, for different values of the ModMax parameter $\gamma$. The black dashed curves correspond to the linear theory. For $\gamma\neq 0$, nonlinear effects steepen the dynamics of the pulse, being this effect more important as $\gamma$ increases. We took $N=8001$ points for a numerical grid of length $L=40$.}
  \label{fields-dynamics}
\end{figure*}

We compute the components of the angular momentum directly from Eq. (\ref{eq:MomentoAngularENL}). Again, imposing the Ansatz (\ref{eq:Ansatz}), we get
\begin{eqnarray}
\label{eq:NumMomentoAngular}
\mathcal{J}_x (t) &=& L^2 \int_{-L/2}^{L/2} \left(z E_z B_x - \frac{L}{2}(E_x B_y - E_y B_x)\right)\mathcal{L}_F \,dz , \nonumber \\
\mathcal{J}_y (t) &=& L^2 \int_{-L/2}^{L/2} \left(z E_z B_y + \frac{L}{2}(E_x B_y - E_y B_x)\right)\mathcal{L}_F \,dz , \nonumber \\
\mathcal{J}_z (t) &=& - \frac{L^3}{2}\int_{-L/2}^{L/2} E_z (B_x + B_y)\,\mathcal{L}_F \,dz \nonumber .
\end{eqnarray}
And so 
\begin{equation} \label{num:MomAngular}
    \mathcal{J}(t) = \sqrt{\mathcal{J}_x^2 (t) + \mathcal{J}_y^2 (t) + \mathcal{J}_z^2 (t)}.
\end{equation}
Notice that as for vacuum configurations we can choose $D_z$ to be always zero, then $E_z = 0$ and thus $\mathcal{J}_z\equiv 0$.

\subsubsection{Birefringence indices and comparison with other nonlinear extensions}

We also keep track of the birefringence indices $\lambda_\pm$ introduced in equation (\ref{biref-indeces}) during the numerical evolution. In ModMax, the conformal invariance immediately implies that $\lambda_- = 0$, being this the case for any conformally invariant nonlinear extension. In fact, we compute numerically both indices from the simulations carried out, and verify that $\lambda_-$ remains very small (roughly of the order of the numerical precision). We exhibit birefringence effects also in other nonlinear models, particularly in Born-Infeld and Gauss-Bonnet Electrodynamics. The former theory also shares conformal invariance, and moreover it does not exhibit birefringence, as $\lambda_-=\lambda_+\equiv 0$. We verify this behavior through numerical simulations, and also compare the effects between Gauss-Bonnet and ModMax electrodynamics.  

\subsection{Results}

\subsubsection{ModMax evolution and Bekenstein inequalities}

Figure \ref{fields-dynamics} displays snapshots of the evolution of the electromagnetic fields at times $t=2.5$ (left column) and $t=10.0$ (right column). The initial configuration is given by Gaussian pulses for $D_x$ and $B_x$, while $D_y$ and $B_y$ are set to zero, as specified before. The black dashed curves correspond to the evolution for the case $\gamma = 0$, equivalent to the linear theory. In this case, a smooth wave propagation is expected, with two characteristic modes of constant amplitude and equal speed: one moving to the right and the other one to the left. Since in this regime the theory is linear, no shocks are expected to form. When $\gamma\neq 0$, nonlinear effects start to dominate the dynamics, as it can be noticed from the continuous colored curves. For our simulations, we consider $\gamma$ to vary from $\gamma = 0$ to $\gamma = 5$. In particular, the initial pulse splits in more modes, and all of them admit different speeds and amplitudes. Faster modes escape the numerical domain quite soon, while the slower ones move at a speed comparable to the linear mode. This may cause the eventual formation of shock profiles; i.e., prominent changes of the values of the fields in quite small regions of space. 
From the initial configuration of $D_x$ and $B_x$ evolved in this example, we can see the development of a much more prominent wave front than in the linear case, occurring soon after the initial time, and at approximately $z\sim 2.7$. This behavior is more notable as $\gamma$ increases. Larger values of this parameter are also possible to simulate if improving the spatial resolution. Additionally, as $\gamma$ increases, there are modes which propagate faster, which imply they reach the numerical boundary quite rapidly. Since periodic boundary conditions are imposed for simplicity, long-time simulations with higher values of $\gamma$ imply enlarging the numerical domain, as well as increasing the resolution for capturing eventual shock-wave fronts in an accurate way. Finally, regarding the evolution of $D_y$ and $B_y$, the prominences that characterize the nonlinear evolution become less pronounced as $\gamma$ increases; i.e., the wave fronts either move with a slower propagation speed than the initial wavefront or remain stationary near $x=0$, with the latter behavior becoming more noticeable for $\gamma \geq 2$. This can be appreciated as a consequence of the total variation diminishing property characterizing our numerical method. This feature allows whetter shock predictions on coarse grids, saving computational time and avoiding spurious oscillations during evolution.

\begin{figure}[h]
  \centering
  \includegraphics[scale=0.6]{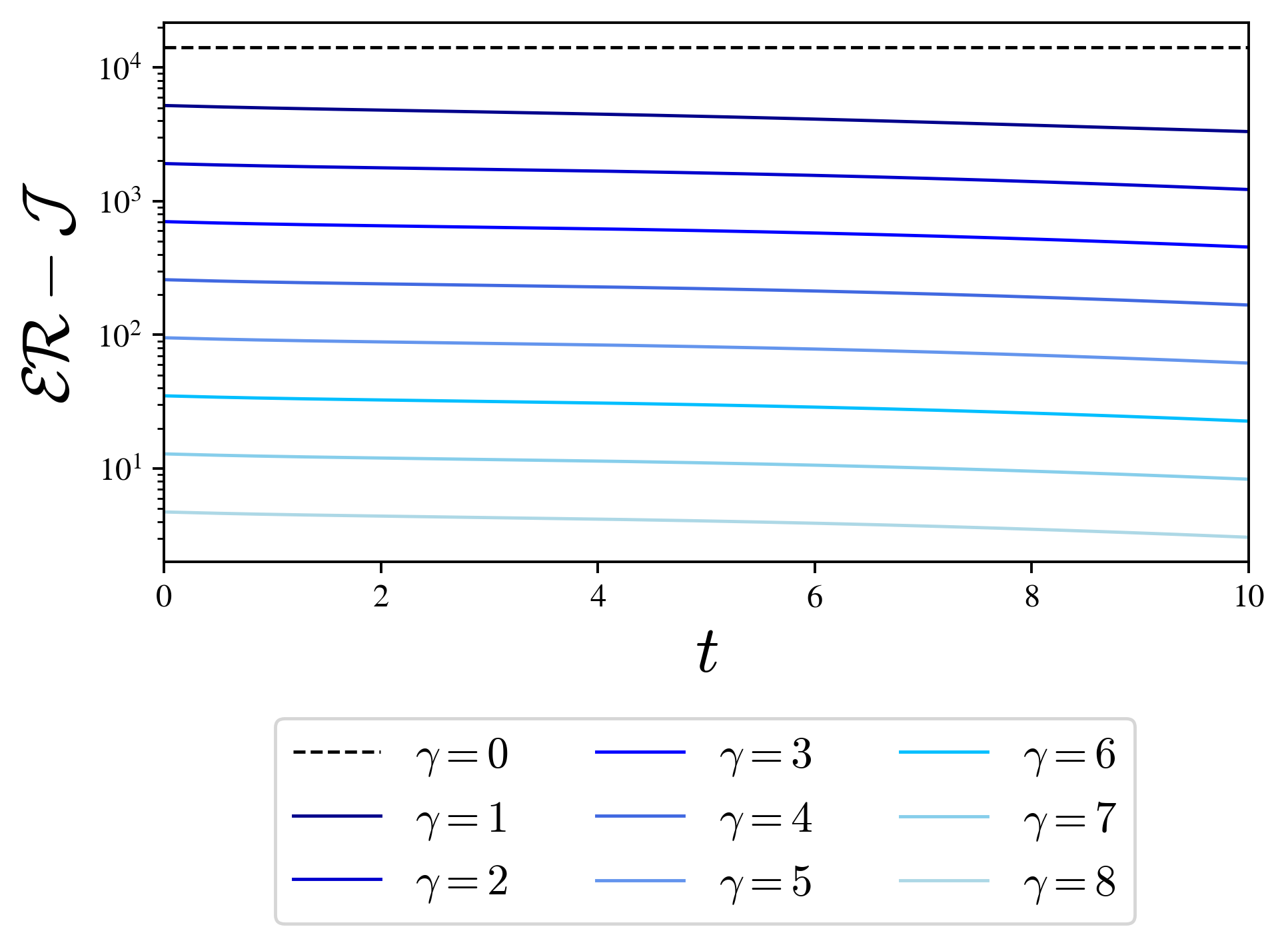}
  \includegraphics[scale=0.6]{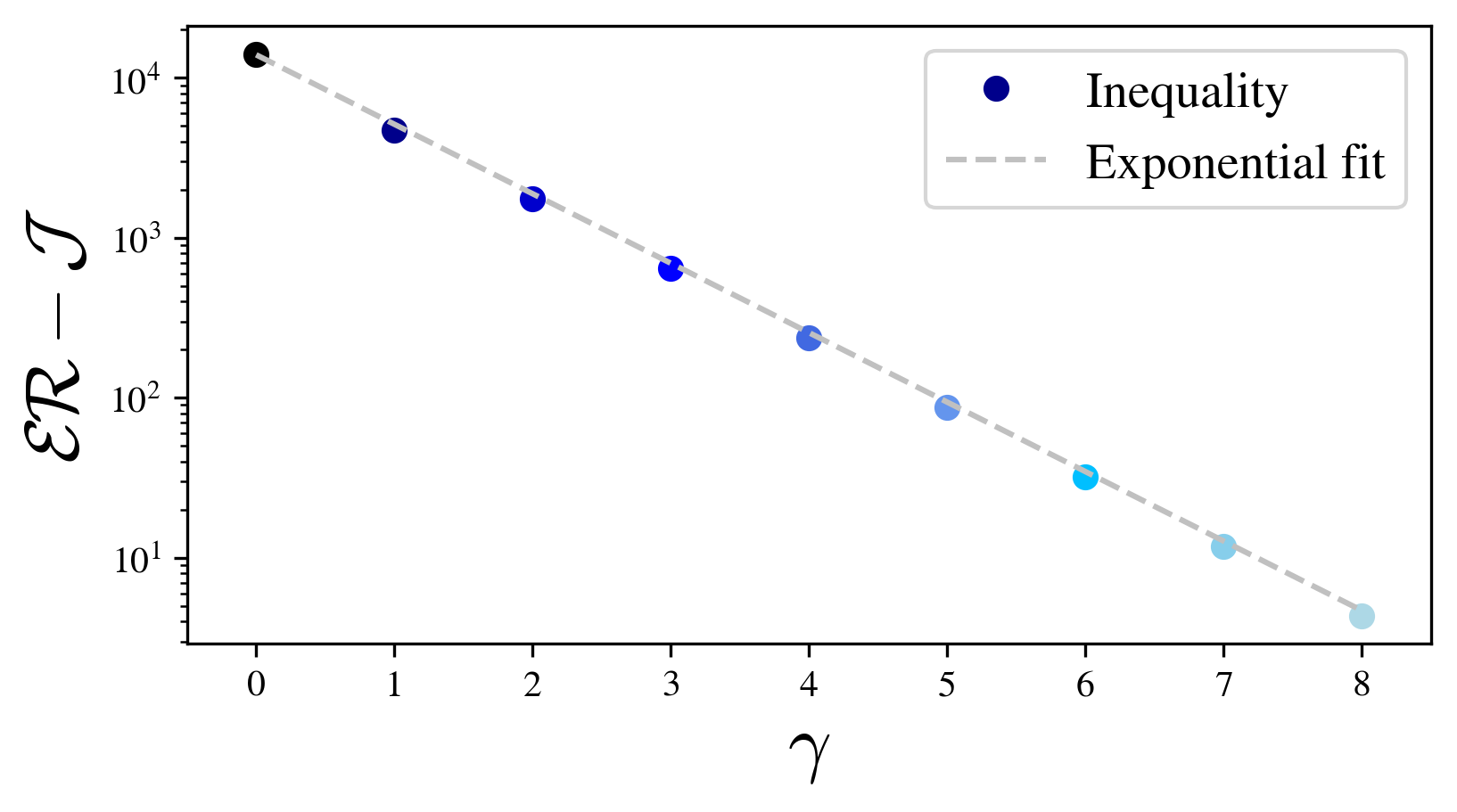}
\caption{\textit{Angular momentum inequality.} \textbf{Top:} Time evolution of the difference $\mathcal{ER}-\mathcal{J}$ for different values of the ModMax parameter $\gamma$. The energy $\mathcal{E}$ and angular momentum $\mathcal{J}$ are computed from the respective formulae (\ref{eq:NumEnergia}) and (\ref{num:MomAngular}), while $\mathcal{R}$ is the radius of the smallest sphere containing the cubic domain. By virtue of Theorem II, a positive value is expected. \textbf{Bottom}: Mean value of the difference $\mathcal{ER}-\mathcal{J}$ as a function of $\gamma$. An exponential scaling is reported (dashed grey line) with fitting determination of $\sim 0.99$.}
  \label{plot-desigualdad}
\end{figure}

We also kept track of physical quantities depending on the dynamical fields; in particular the electromagnetic energy and angular momentum, as computed according to formulae (\ref{eq:NumEnergia}) and (\ref{num:MomAngular}), respectively. We evaluated the inequality (\ref{eq:DesigualdadSinQ}), and confirmed its validity during the whole evolution. Moreover, we performed evolutions of a large class of initial data configurations, for which the inequality was also satisfied. Figure \ref{plot-desigualdad} displays the difference $\mathcal{ER}-\mathcal{J}$, for which a positive value is expected. As time evolves, and for each $\gamma$, the value of the difference remains practically constant, although a small decrease is expected due to the dissipation of energy caused from shock formation. This can be seen from the top panel of Figure \ref{plot-desigualdad}. We report that the value of the difference $\mathcal{ER}-\mathcal{J}$ decreases exponentially as $\gamma$ increases, as shown in the bottom panel. For each value of $\gamma$, we plotted the mean value of the difference, and compared the trend with an exponential fit, getting a fitting determination of $\sim 0.99$, indicating that the exponential aligns with the data almost perfectly.

\begin{figure}[h]
  \centering
  \includegraphics[scale=0.6]{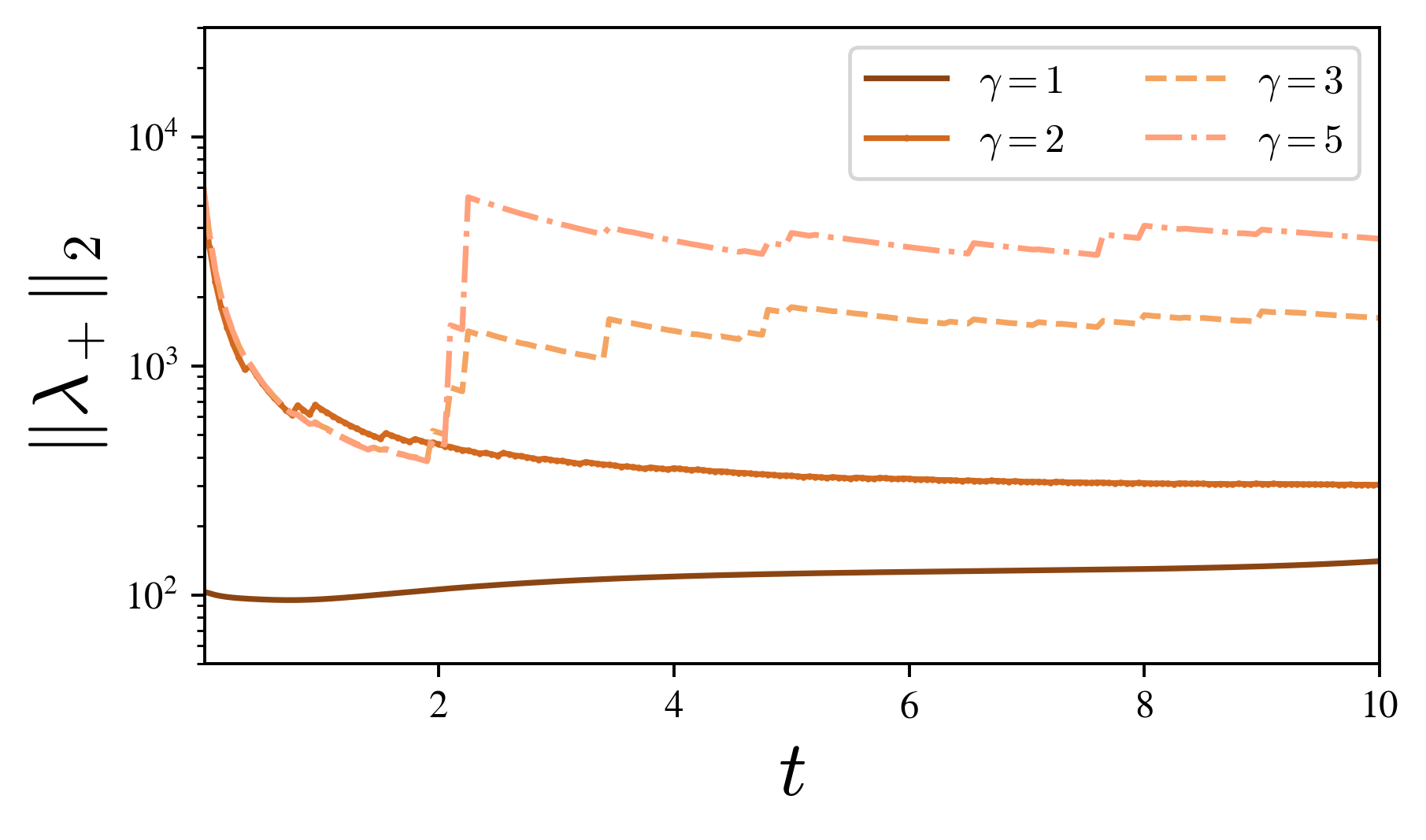}
  \includegraphics[scale=0.6]{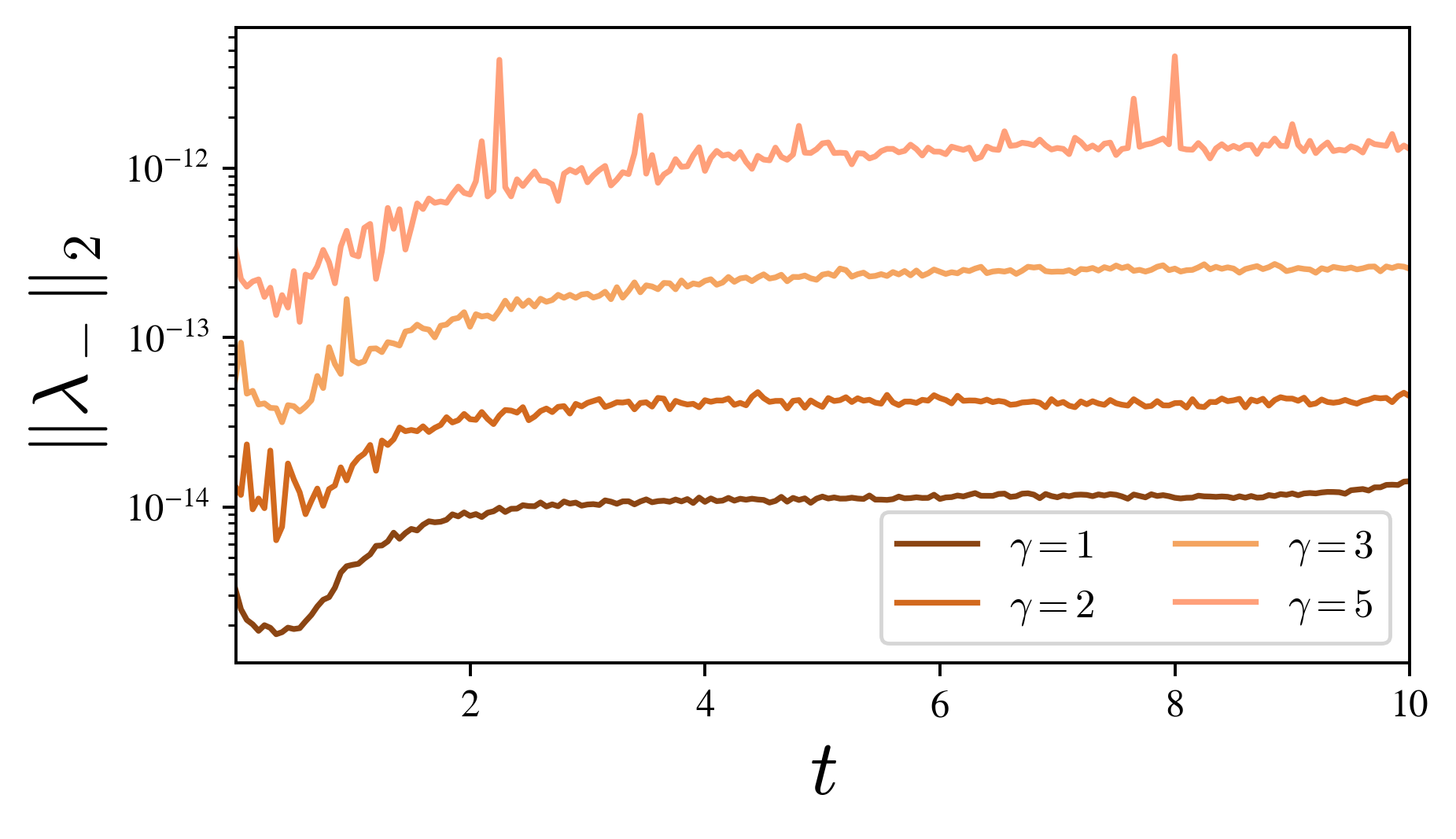}
\caption{\textit{Birefringence in ModMax.} $L^2$-norm of the birefringence indices $\lambda_+$ (top) and $\lambda_-$ (bottom) as a function of evolution time, for different values of the ModMax parameter $\gamma$. As expected, $\lambda_-$ remains almost zero, while $\lambda_+$ increases notably with $\gamma$, displaying a jump for higher values of $\gamma$, in accordance with a similar behavior observed for the fields at the same evolution time.}
  \label{plot-biref-modmax}
\end{figure}

\begin{figure*}
  \centering
  \includegraphics[scale=0.6]{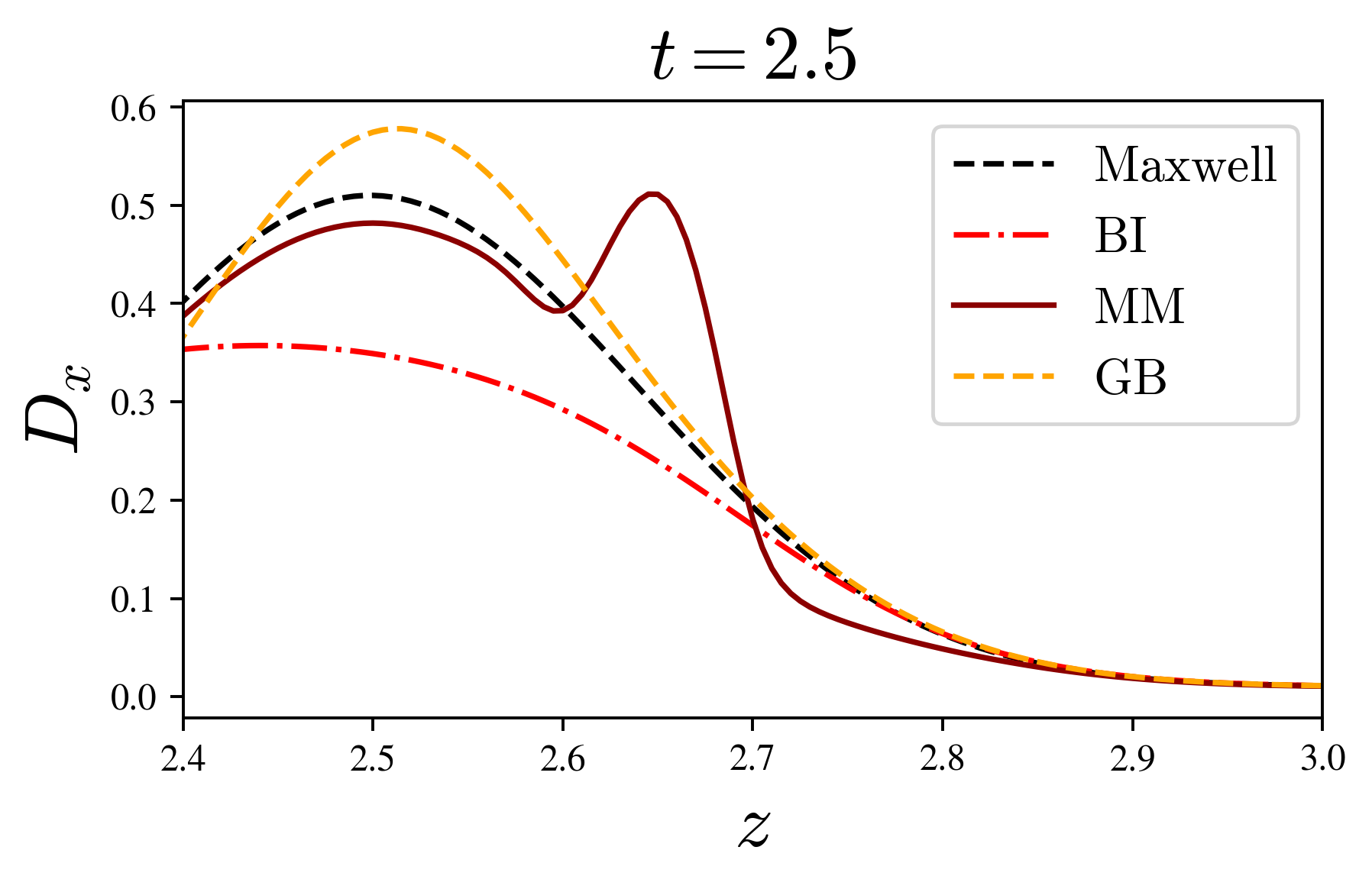}
  \includegraphics[scale=0.6]{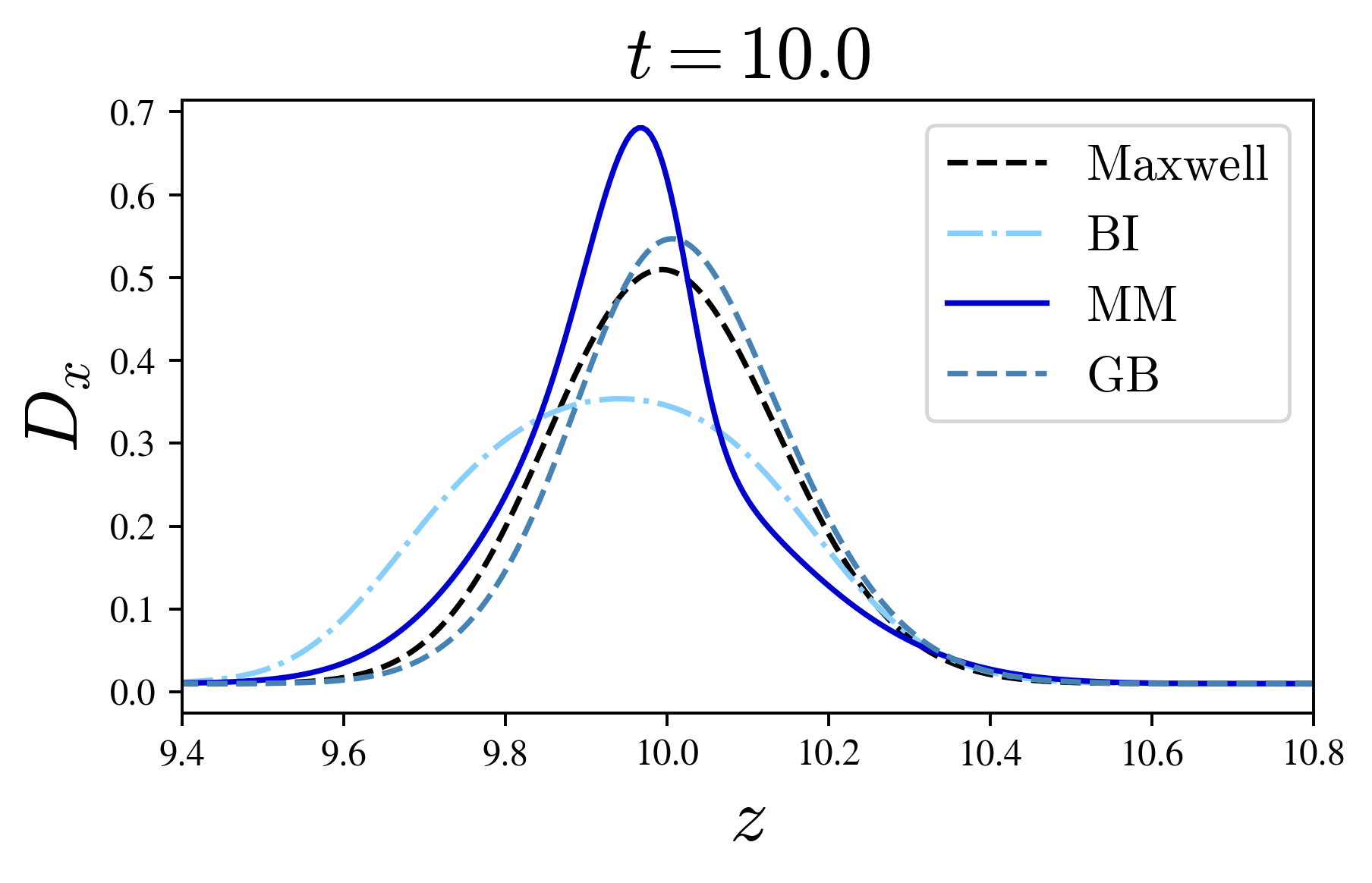}
  \includegraphics[scale=0.6]{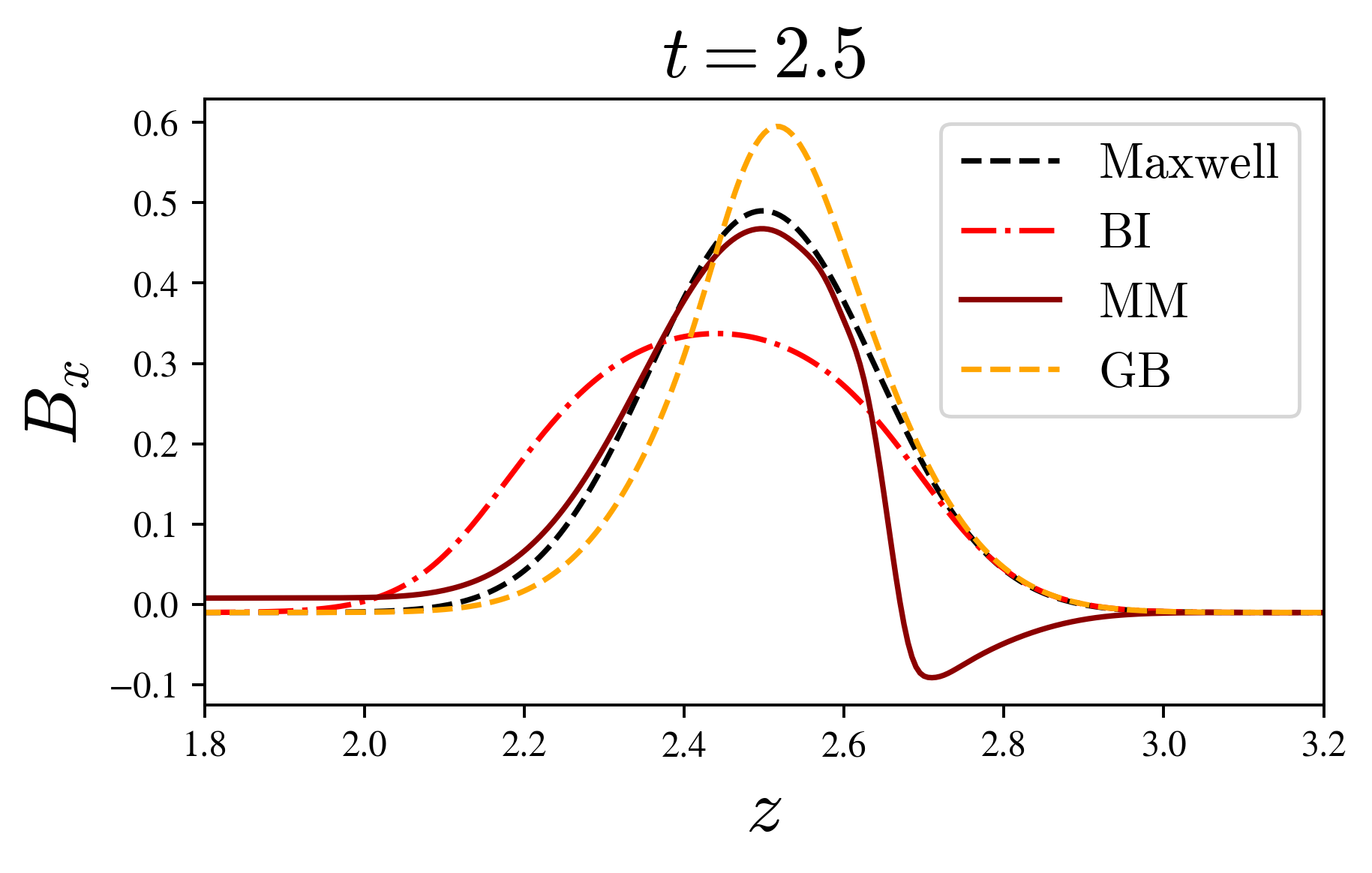}
  \includegraphics[scale=0.6]{comp_Dx_10.0.png}
  \caption{\textit{Dynamics comparison with other nonlinear theories.} Profile snapshots of the evolution carried out in Maxwell (dotted black), Born-Infeld (dotted red), ModMax (solid bordeaux) and Gauss-Bonnet (dotted yellow), at times $t=2.5$ (left column) and $t=10.0$ (right column), with respective coupling constants $T = 0.02$, $\gamma = 5$ and $\alpha = 0.1$. For this configuration of the fields, nonlinear effects are stronger in ModMax, and profiles with lower peaks are wider. For GB and BI, nonlinearities are manifested by smooth deformations of the gaussian pulse with respect to Maxwell evolution.}
  \label{comparison-evolution}
\end{figure*}

\begin{figure}[h]
  \centering
  \includegraphics[scale=0.6]{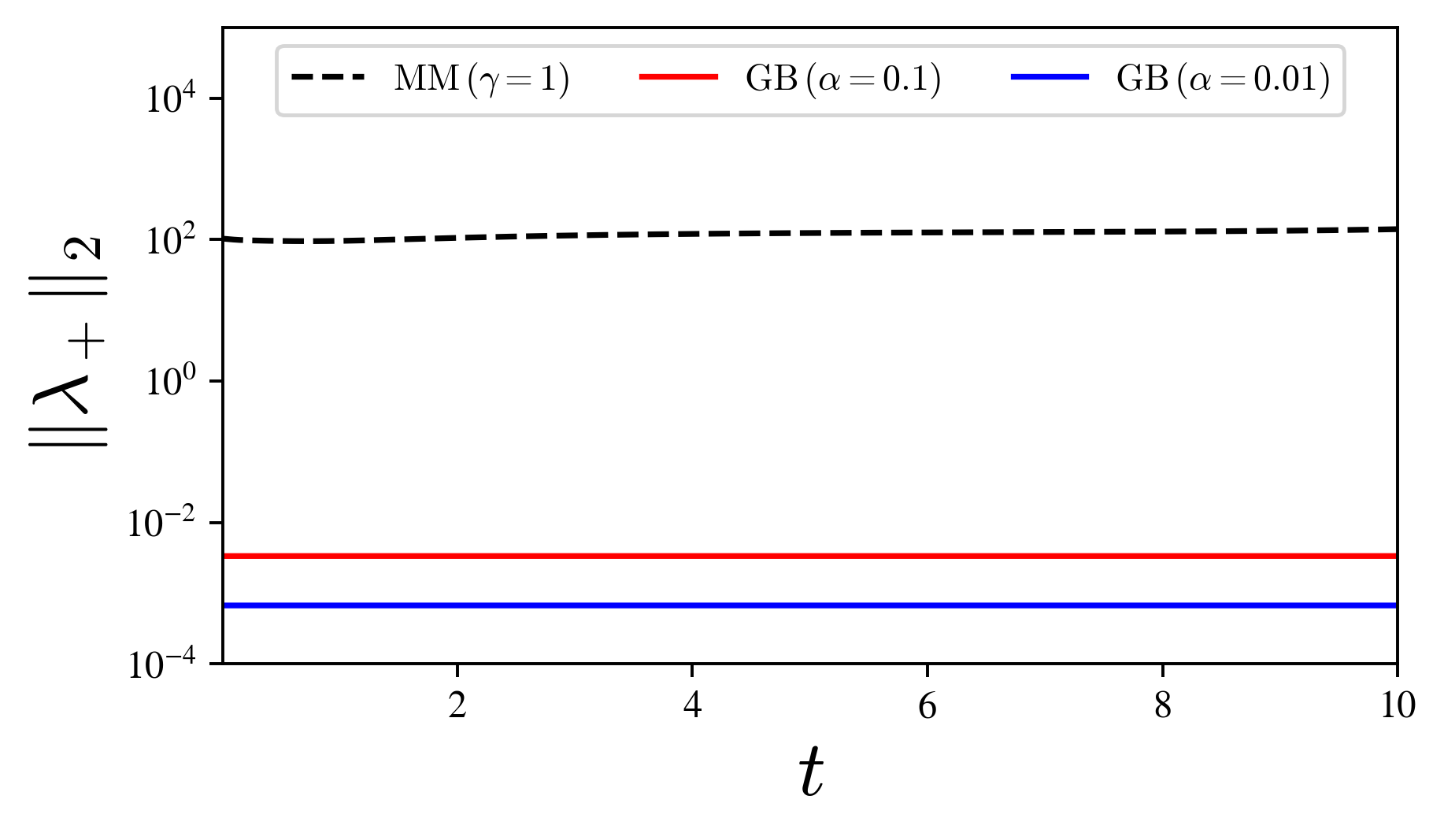}
\caption{\textit{Birefringence in other nonlinear theories.} Comparison of the $L^2$ norm of $\lambda_+$ as a function of time between ModMax evolution with $\gamma=1$ (dotted black) and Gauss-Bonnet dynamics with couplings $\alpha=0.1$ (solid red) and $\alpha=0.01$ (solid blue).}
\label{plot-comp-bireff}
\end{figure}

\subsubsection{Birefringence}

\begin{figure}[h]
  %\centering
  \includegraphics[scale=0.6]{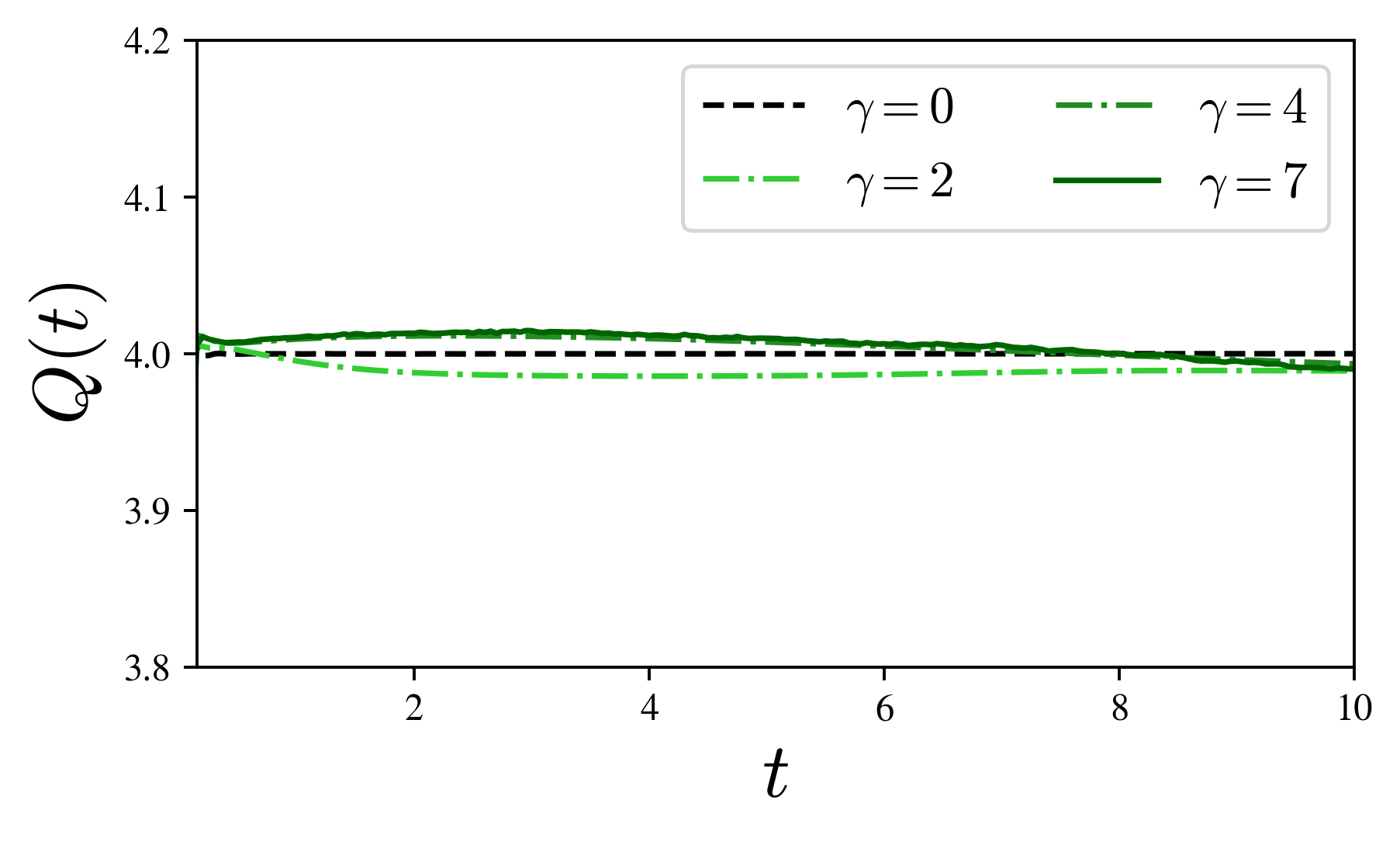}\\
  \includegraphics[scale=0.6]{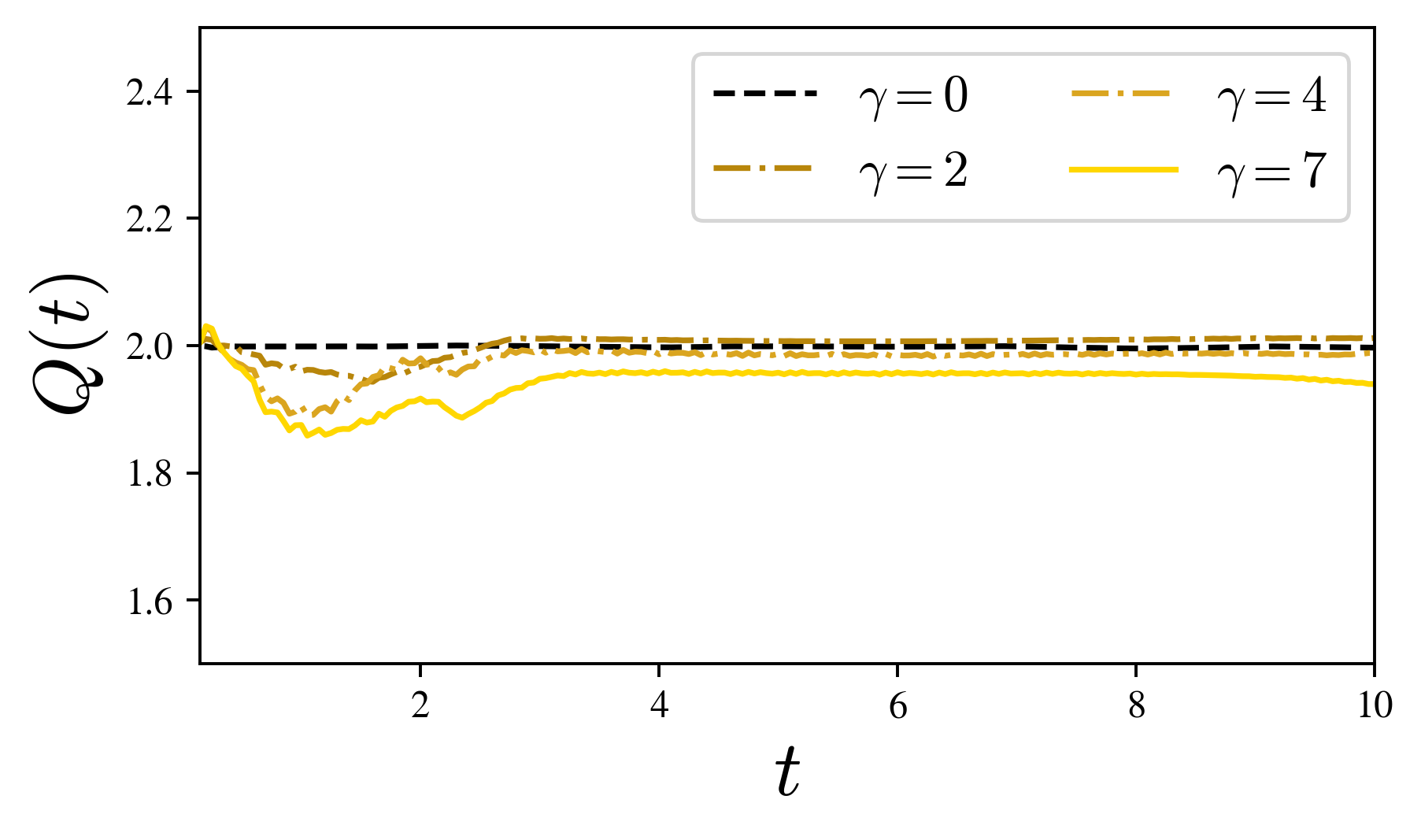}
\caption{\textit{Convergence tests.} Order of accuracy in time (top) and space (bottom) of our numerical scheme, for different values of the ModMax parameter $\gamma$. All the simulations where performed from the initial data specified in Table \ref{Tabla}.  For time convergence, we fixed $N=8001$ and set the time steps as $\Delta t = 0.005$ (low resolution), $\Delta t/2$ (medium resolution) and $\Delta t/4$ (high resolution). Instead, for testing convergence in space, a time step of $\Delta t = 0.001$ was fixed, while the spatial resolutions where $N_1 = 4001$ (low), $N_2 = 8001$  (mid) and $N_3 = 16001$ (high). Derivatives where discretized using second-order finite difference operators which satisfy the summation by parts property.}
  \label{convergence-plots}
\end{figure}

We now report on the results obtained for the birefringence in ModMax, from the numerical simulations shown in the previous plot. Figure \ref{plot-biref-modmax} shows the $L^2$-norm of the birefringence indices $\lambda_+$ and $\lambda_-$ introduced in Eq. (\ref{plot-comp-bireff}). As expected from the conformal invariance exhibited by the theory, there is one mode (in this case the one represented by $\lambda_-$) which remains very close to zero, giving rise to the dispersion relation of the pure wave equation in flat space, i.e. $\omega^2 - k^2 = 0$. On the other hand, the norm of $
\lambda_+$ is quite greater than zero, indicating an important birefringence effect. In fact, their value scales with the parameter $\gamma$, and the effect gets more and more notable as non-linearities become important. Also, for higher values of $\gamma$, the behavior of $\lambda_+$ is comparable to the one for lower values until $t\sim 2.5$, in which the evolution displays a jump in the value of the fields. This effect is still more evident for higher values of $\gamma$.

Although the presence of birefringence is expected for nonlinear theories of electromagnetism, the results found for ModMax are still remarkable, as the theory (even though nonlinear) shares all the symmetries of Maxwell's theory, and in principle one might not expect birefringence (like in Maxwell), which would be suggested by the vanishing of $\lambda_-$. On the contrary, not only $\lambda_+$ is different from zero, but also it actually increases in norm in highly-nonlinear regimes. This shows a significant difference with respect to other nonlinear extensions of Electromagnetism, which do not even present birefringence. This contrast will be explicitly shown in what follows.

\subsubsection{Comparison with other NLED theories}

For the sake of comparison, we also evolved two other nonlinear theories for electromagnetism, analyzing their dynamics and computing birefringence effects. The first one is \textit{Born-Infeld} (BI) theory \citep{Born:1934gh}, whose lagrangian density reads
    \begin{equation}
        \mathcal{L}_{\text{\scriptsize{BI}}} = -T^2\sqrt{-\det\left(\eta_{\mu\nu} +\frac{F_{\mu\nu}}{T}\right)}+T^2.
    \end{equation}
    The explicit evolution equations for the numerical implementation are obtained from the corresponding Hamiltonian, which reads
\begin{eqnarray}        \mathcal{H}_{\text{\scriptsize{BI}}}&=&T^2\sqrt{1+\frac{1}{T^2}\left(|\mathbf{D}|^2+|\mathbf{B}|^2\right)+\frac{1}{T^4}|\mathbf{D}\times\mathbf{B}|^2} - T^2. \nonumber \\
    &&
\end{eqnarray}

Here, $T>0$ is a constant with dimensions of energy density. While it represents the maximum possible value of the electric field, it also has an interpretation as (the square root of) a 3-brane tension in string-theory. Maxwell electrodynamics is recovered in the ``weak field'' limit, i.e. when $T\to\infty$ \citep{Russo:2022qvz}. Also, the theory is duality invariant (as Maxwell and ModMax). Instead, the ``strong-field'' limit ($T\to 0$) the theory gives rise to Bialynicki-Birula electrodynamics \citep{PhysRev.130.465}, which has an $SL(2,\mathbb{R})$ duality invariance.

We also studied the evolution in \textit{Gauss-Bonnet} (GB) Electrodynamics, with lagrangian density
    \begin{equation} \label{GB-lagrangian}
        \mathcal{L}_{\text{\scriptsize{GB}}} = -F - \frac{3\alpha}{2}G^2.
    \end{equation}
This theory is a subclass of the Kaluza-Klein family of nonlinear models for Electromagnetism, whose lagrangian is obtained by adding a Gauss-Bonnet term to the Einstein-Hilbert action for General Relativity in five spacetime dimensions \citep{Kaluza:1921tu, Klein:1926tv}. In string-theory, the lagrangian (\ref{GB-lagrangian}) is interpreted as a first-order correction to Einstein's gravity, and the parameter $\alpha$ accounts for a small perturbation that depends on fundamental physical quantities (as the electron's charge and mass, the speed of light and the Planck's constant). For our numerical implementation, we found more useful to compute the constitutive relations directly for the lagrangian formulation. For doing so, we computed the permittivities $\varepsilon = -\mathcal{L}_F$ and $\nu = \mathcal{L}_G$, and from them obtaining the constitutive relations from 
\begin{eqnarray*}
    \textbf{E} &=& \frac{1}{\varepsilon}\left(\mathbf{D}-\nu\mathbf{B}\right)\\
    \textbf{H} &=& -\frac{\nu}{\varepsilon}\mathbf{D} + \left(\varepsilon+\frac{\nu^2}{\varepsilon}\right)\mathbf{B}.
\end{eqnarray*}
With all this, we wrote the explicit evolution equations in the form given by (\ref{eq:EcEvolucion}, \ref{eq:EcEvolucion2}).

In order to strength out the nonlinear effects within the regime of validity of both theories, we performed numerical evolutions taking $T=0.02$ and $\alpha=0.1$. Figure \ref{comparison-evolution} displays snapshots comparing the evolution of ModMax with the nonlinear dynamics yield by both BI (dotted red) and GB (dotted yellow), starting from the same initial data as before. At $t=2.5$, the evolution profiles of BI and GB remain quite smooth, unlike the ModMax case, which exhibits a pronounced change in the profile of both $D_x$ and $B_x$ fields. This indicates that nonlinear effects are stronger in ModMax, while the relative amplitude of the peaks are quite comparable. In particular, profiles with lower peaks are wider, a consequence of being governed by evolution equations obeying conservation laws (even in the highly-nonlinear regime). For $t=10$, we see that this configuration for the fields does not seem to develop shocks or discontinuities at the same time scale as ModMax does, and the nonlinearities are manifested simply by small deformations of the gaussian pulse with respect to the profile yield from Maxwell evolution.

Finally, we compared the birefringence effects resulted from the above simulations. As it is known, BI does not exhibit birefringence, unlike GB and ModMax. The comparison of the order of magnitudes of the norm of $\lambda_+$ is displayed in figure \ref{plot-comp-bireff}. The black dotted line corresponds to ModMax evolution with $\gamma = 1$, while read and blue lines correspond to GB evolution with two different values of the coupling constant: $\alpha=0.01$ (blue) and $\alpha=0.1$ (red). As expected, for smaller $\alpha$ the theories approaches to Maxwell electrodynamics, which does not exhibit birefringence. Moreover, increasing $\alpha$ by one order of magnitude still yields a quite small value for the norm of $\lambda_+$ when compared with the ModMax evolution (even with one of the smallest values of $\gamma$ considered for the evolution). This seems to indicate that the birefringence effect is visible as a consequence of the nonlinearity of the theory, but it is stronger when shocks/discontinuities are developed during evolution.

\subsubsection{Convergence tests}

In order to validate the robustness of our results, we performed convergence tests in space and time. As it was pointed out in the previous section, we implemented a fourth-order Runge-Kutta scheme for the time integration, adding artificial viscosity for avoiding spurious oscillations near eventual shock fronts. For the computation of spatial derivatives, we implemented second-order accurate finite difference operators, which satisfy the property of summation by parts. This allows semi-discrete energy estimates for the corresponding initial-boundary value problem, guaranteeing numerical stability during the whole evolution.
We assessed the convergence and stability of the scheme for each value of the parameter $\gamma$. For the time convergence, we fixed the number of grid points to $N=8001$ in a numerical domain from $z_i = -20$ to $z_f=20$. We performed three runs with time steps $\Delta t = 0.005$ (low resolution), $\Delta t/2$ (medium resolution) and $\Delta t/4$ (high resolution), and computed the precision coefficient \cite{KreissOrtiz2014}
\begin{equation}
\label{eq:CoePrecision}
Q(t)=\log_2\,\frac{||U_{\mbox{\tiny{low}}} - U_{\mbox{\tiny{med}}}||}{||U_{\mbox{\tiny{med}}} - U_{\mbox{\tiny{high}}}||}\;,
\end{equation}
where $U$ is the evolution field, and $||U_{\mbox{\tiny{a}}} - U_{\mbox{\tiny{b}}}|| = \sum_i |U_{\mbox{\tiny{a}},i} - U_{\mbox{\tiny{b}},i}|$ is the $L^1$ norm of the
difference of the two discretized fields over all the spatial grid $i=1,\cdots,N$. We compared the accuracy order obtained from (\ref{eq:CoePrecision}) with the expected nominal value, which in this case should be $\sim 4$. As it can be seen from the top panel of Figure \ref{convergence-plots}, our simulations converge with the expected rate, even for larger values of $\gamma$. The spatial convergence is displayed in the bottom panel of Figure \ref{convergence-plots}. In this second test, we fixed the time step to be $\Delta t = 0.001$, considered three different spatial resolutions: $N_1 = 4001$ (low), $N_2 = 8001$ (mid) and $N_3 = 16001$ (high), and computed the coefficient (\ref{eq:CoePrecision}) as before. In this case, the result should be the order of accuracy of the finite difference operators used, which in our case is $2$. As expected, our simulations display an approximately second order convergence in space, ensuring a correct behavior of the numerical scheme even for large values of $\gamma$. This assessment allows us to trust in the robustness of our results.

\section{Conclusions}
\label{sec-conclusions}
In this article we inspected peculiar features about the dynamics of the ModMax family of nonlinear extensions of Electromagnetism. This theory turned out to be remarkable for it is the only one that preserves the conformal and duality symmetries characterizing Maxwell’s electrodynamics. By looking at the characteristic structure of the theory in a purely covariant way, we used a simple geometric criterion for the hyperbolicity of nonlinear theories of Electromagnetism and showed that ModMax is symmetric-hyperbolic. This property implies that it admits a well-posed Cauchy Problem, which in turn allows to conduct stable numerical simulations.

We also explored a series of geometric inequalities in ModMax, in the context of Bekenstein bounds. In particular, we proved three inequalities relating electromagnetic energy, angular momentum, charge and a measure of size in bounded regions of space. Of course, the validity of such inequalities depends on the measure of size, and they might actually fail for some cases. Nevertheless, the ones shown here support the validity of the original universal relations conjectured by Bekenstein, which represented a sound effort for solving Black Hole Thermodynamics. Thus, since these relations are supposed to have universal validity, it is reasonable to use them for assessing the feasibility of NLED candidates. In particular, the inequality (\ref{eq:DesigualdadSinQ}) shown for ModMax, implies that the electromagnetic energy of the system has to be greater or equal to its rotational kinetic energy, with a ratio depending on the speed of light and the radius of the smallest sphere containing the system. Thus, this condition seems to be related with the causal structure of the theory. It would be of interest a study about the rigidity of these inequalities in ModMax, and even in more general NLED theories.

Finally, we conducted numerical simulations of the ModMax system of equations, in 1+1 dimensions. The set of evolution equations constitutes a system of nonlinear conservation laws, and the corresponding fluxes can be written as algebraic functions of the dynamical variables. We evolved the equations in a simple way, by adding artificial viscosity, allowing stable formation and propagation of shocks. We applied our numerical evolution to verify the inequality between energy and angular momentum, as well as to compute birefringence effects, comparing their strength with other nonlinear theories in similar regimes. As a future perspective, we are interested in extending our implementation to full 3D simulations, for which alternative high-resolution shock-capturing schemes would be fundamental in potential applications to black hole solutions and astrophysical phenomena.

%%%%%%%%%%%%%%%%%%%%%%%%%%%%%%%%%%%%%%%%%%

\section*{Acknowledgements}
We would like to thank Enrico Barausse, Miguel Bezares, Luis Lehner and Oscar Reula for discussions throughout the realization of this work. We are also grateful to Alejandro Perez for reading a first version of the manuscript, and to the referee for suggesting the numerical study of birefringence in ModMax. MR acknowledges support from the European Union’s H2020 ERC Consolidator Grant ``GRavity from Astrophysical to Microscopic Scales'' (Grant No. GRAMS-815673), the PRIN 2022 grant ``GUVIRP - Gravity tests in the UltraViolet and InfraRed with Pulsar timing'', the EU Horizon 2020 Research and Innovation Programme under the Marie Sklodowska-Curie Grant Agreement No. 101007855 and the MUR PRIN Grant No. 2022-Z9X4XS funded by the European Union (Next Generation EU).
\\
\\
\textit{This work is dedicated to the memory of Prof. Sergio Dain, an exceptional inspiration for a generation of scientists.}

\appendix

\section{A counterexample of the original energy-charge inequality}
\label{app-counterex}

In this Appendix, we give a counterexample of the standard inequality between energy, charge and size in ModMax Electrodynamics, which reads
\begin{equation}
\mathcal{E} \geq \frac{Q^2}{8 \pi \mathcal{R}}.
\label{eq:E-Q-vieja}
\end{equation}
To demonstrate that this version of the inequality is not valid in ModMax, it is important to notice a rigidity condition shown by Dain in \cite{Dain2015}: the equality in (\ref{eq:E-Q-vieja}) is only valid for the electrostatic configuration given by a thin spherical shell with a constant surface charge density and radius $\mathcal{R}$. Thus, to find a counterexample, it suffices to show that the energy density $u$ of ModMax can be less than the energy density $u_M$ of Maxwell's theory when the rigidity condition holds. In other words, we should see that there exists some field configuration such that $u < u_M^e$, where the superscript $e$ refers to the electrostatic configuration (for which the equality holds). 

Recalling the expressions for the energy densities from (\ref{eq:uModMax}) and (\ref{eq:uMaxwell}), the difference $\Delta u := u - u^e_M$ results in
\begin{equation}
\Delta u=\frac{\textbf{E}^2}{2}(\cosh(\gamma)-1)+\frac{\textbf{B}^2}{2}\left(\cosh(\gamma)-\frac{\textbf{B}^2}{\sqrt{\eta}}\sinh(\gamma)\right), \nonumber
\end{equation}
where
\begin{equation}
\eta=(\textbf{E}^2-\textbf{B}^2)^2+4\textbf{E}^2 \textbf{B}^2 \cos^2(\alpha),\nonumber
\end{equation}
and $\alpha$ is the angle between the two fields. Now, if $\alpha=\pi/2$ and $\textbf{B}^2 > \textbf{E}^2$, and multiplying by $\textbf{B}^2-\textbf{E}^2$,  we get 
\begin{equation}
(\Delta u)(\textbf{B}^2-\textbf{E}^2)=\frac{1}{2}\left(\textbf{B}^4 e^{-\gamma} - \textbf{E}^4 \cosh(\gamma) - \textbf{E}^2\textbf{B}^2 + \textbf{E}^4\right)\nonumber
\end{equation}
In turn, take $\textbf{B}^2$ as $\textbf{B}^2 = z \textbf{E}^2$, where $z>1$ (such that $\textbf{B}^2 > \textbf{E}^2$). Thus, we get that the sign of $\Delta u$ is
\begin{equation}
\text{sgn}(\Delta u)=\text{sgn}\left[z^2 e^{-\gamma} - z + (1-\cosh(\gamma))\right].\nonumber
\end{equation}

We then aim to demonstrate that, for a given $\gamma$, there exists a value of $z$ such that the function $f(z) = z^2 e^{-\gamma} - z + (1 - \cosh(\gamma))$ is negative. This condition is equivalent to showing that the roots of $f(z)$ are real and distinct (as $z > 1$). In fact, the roots $z_+$ and $z_-$ of $f(z)$ are given by
\begin{eqnarray}
z_+&=&\frac{1+\sqrt{1-4 e^{-\gamma}(1-\cosh(\gamma))}}{2 e^{-\gamma}}, \nonumber\\
z_-&=&\frac{1-\sqrt{1-4 e^{-\gamma}(1-\cosh(\gamma))}}{2 e^{-\gamma}}\nonumber
\end{eqnarray}
In particular,
\begin{equation}
z_+\geq e^{\gamma} \geq 1,\nonumber
\end{equation}
concluding that $\Delta u < 0$ in the interval $1 < z < z_+$ for all $\gamma$, contradicting Dain's rigidity result.

%\clearpage	
%% \newpage

\bibliographystyle{ieeetr}
\bibliography{main}

\end{document}